\documentclass[twocolumn,aps,prd,preprintnumbers,superscriptaddress,nofootinbib]{revtex4-1}
\usepackage{amsmath,amssymb,amsfonts,mathrsfs}
\usepackage[dvipdfmx]{graphicx}
\usepackage{enumerate} 
\usepackage{colordvi} 
\usepackage{color} 
\usepackage{bm}
\usepackage{epsf,color}
\usepackage{array}
\usepackage{siunitx}
\usepackage{float}

\begin{document}
\title{Impact of correlated magnetic noise on the detection of stochastic 
gravitational waves: Estimation based on a simple analytical model}
\author{Yoshiaki Himemoto}
\affiliation{Department of Liberal Arts and Basic Sciences, College of Industrial Technology, Nihon University, Narashino, Chiba 275-8576, Japan}
\author{Atsushi Taruya}
\affiliation{Center for Gravitational Physics, Yukawa Institute for Theoretical Physics, Kyoto University, Kyoto 606-8502, Japan}
\affiliation{
Kavli Institute for the Physics and Mathematics of the Universe, Todai Institutes for Advanced Study, the University of Tokyo, Kashiwa, Chiba 277-8583, Japan (Kavli IPMU, WPI)}
%
%
%
%
%
\date{\today}
\begin{abstract}
After the first direct detection of gravitational waves (GW), detection of stochastic background of GWs is an important next step, and the first GW event suggests that it is within the reach of the second-generation ground-based GW detectors. Such a GW signal is typically tiny, and can be detected by cross-correlating the data from two spatially separated detectors if the detector noise is uncorrelated. It has been advocated, however, that the global magnetic fields in the Earth-ionosphere cavity produce the environmental disturbances at low-frequency bands, known as Schumann resonances, which potentially couple with GW detectors. In this paper, we present a simple analytical model to estimate its impact on the detection of stochastic GWs. The model crucially depends on the geometry of the detector pair through the directional coupling, and we investigate the basic properties of the correlated magnetic noise based on the analytic expressions. The model reproduces the major trend of the recently measured global correlation between the GW detectors via magnetometer. The estimated values of the impact of correlated noise also match those obtained from the measurement. Finally, we give an implication to the detection of stochastic GWs including upcoming detectors, KAGRA and LIGO India. 
The model suggests that LIGO Hanford-Virgo and Virgo-KAGRA pairs are possibly less sensitive to the correlated noise, and can achieve a better sensitivity to the stochastic GW signal in the most pessimistic case.
\end{abstract}

\preprint{YITP-17-29}
\maketitle


\section{Introduction}
\label{sec:intro}
%
%
The detection and measurement of a stochastic background of gravitational waves (GWs) is one of the most exciting challenges among the observations of various GW sources. Stochastic GWs are produced by an incoherent superposition of an extremely large number of GWs from the unresolved astrophysical sources and/or high-energy cosmological phenomena such as inflation, cosmic strings, and phase transitions (For a review, see e.g., Ref.~\cite{Maggiore:1999vm}). In particular, the first direct detection of the GW event by the LIGO detectors \cite{Abbott:2016blz} indicates that the expected number of binary black holes like GW150914 is huge, and the sum of their GWs can be viewed as a stochastic background whose amplitude is within the reach of the second-generation GW detectors \cite{TheLIGOScientific:2016wyq} (see also \cite{1674-4527-11-4-001,Zhu:2011bd, Rosado:2011kv,Marassi:2011si,Wu:2011ac,Wu:2013xfa, Mandic:2016lcn}). 
The detection of such a GW signal would, thus, give many hints and clues to clarify the formation and evolution of cosmological binary black holes, complementary to the detection of individual GW events. Further, there might also be stochastic GWs from the cosmological origin at the frequencies relevant for the ground-based detectors, and if detected, a huge impact on cosmological physics in the early Universe is expected.

In general, a detection of such a stochastic signal is very difficult only with a single GW detector. This is because the signal is basically very tiny, and resembles detector's noise. To discriminate between stochastic GW and detector's noise, a standard technique is to use the output data from the multiple set of detector, and to take the cross correlation between them. Since GWs induce the nonvanishing correlation between the two data streams, the cross-correlation technique offers a unique way to isolate the stochastic GW signals, if detector's noises are totally uncorrelated. The validity of this assumption, however, needs to be carefully scrutinized.

Refs.\cite{Allen:1997ad, Christensen:1992wi} have pointed out the possibility that the global disturbances arising from the (stationary) electromagnetic fields on the Earth, known as Schumann resonances \cite{1952ZNatA...7..149S,1952ZNatA...7..250S}, produce the noise correlation through the coupling with magnets used in the laser interferometers system (see also \cite{Kowalska-Leszczynska:2016low} for short duration magnetic field transients). Later, using the magnetometers, the Schumann resonances have been measured at LIGO Hanford/Livingston \cite{Harry:2010zz,TheLIGOScientific:2014jea} and Virgo \cite{TheVirgo:2014hva} sites, and a significant correlation has been found \cite{2013PhRvD..87l3009T}. While the impact on the detection of stochastic GW, inferred from the measured amplitude of Schumann resonances, turned out to be less significant for the first-generation detectors, the advanced second-generation detectors which substantially improves the detector sensitivity can reach the expected amplitude of the correlated magnetic noises, and it is now a great concern. Subtraction and mitigation of the correlated noise is therefore very important issue, and combining the geophysical data monitored outside the GW detectors, a more efficient methodology needs to be developed, together with the experimental efforts on reducing the coupling with magnetic fields (see e.g., Refs.~\cite{2014PhRvD..90b3013T, Coughlin:2016vor} along this direction).

On the other hand, the properties of the correlated magnetic noises are not yet fully understood, and we think that the understanding of it still deserves a further investigation from both the qualitative and quantitative points of view. One example are the magnetic noise spectra measured by Ref.~\cite{2013PhRvD..87l3009T}, which exhibit a notable difference between pairs of GW detectors. This suggests that the geometrical setup of the detector pair may play an important role, and it may provide a clue to exploit an efficient mitigation method for the correlated magnetic noises. In this paper, we present a simple model of correlated noises, and study the role of the geometric configuration on the correlated magnetic noises.  In particular, we derive the analytic expressions for correlated noise spectrum, whose properties are determined not only by the magnetic field spectrum but also by the geometry of the detector pair, together with the function characterizing the coupling between detectors and magnetic fields. While the analytic model is constructed based on several simplification and assumptions, we will show that the model reproduces the major trend of the measured results by Ref.~\cite{2013PhRvD..87l3009T}. Further, the estimated impact on the detection of stochastic GWs, quantified by the signal-to-noise ratios of the correlated noise [see Eq.(\ref{eq:def_SNR_B})], matches pretty well the result obtained by Ref.~\cite{2014PhRvD..90b3013T}.

The structure of this paper is as follows. In Sec.~\ref{sec:formulation}, we begin by briefly reviewing the cross-correlation analysis to detect stochastic GWs, and discuss the impact of correlated magnetic noise. We then present a simple analytic model of correlated noise in Sec.~\ref{sec:analytic_model}. Despite its simplification, the analytic model possesses several interesting properties, which we will study in detail based on the analytic expressions.  In Sec.~\ref{sec:comparison}, the predictions based on the analytic model are compared to the results in Ref.~\cite{2013PhRvD..87l3009T, 2014PhRvD..90b3013T}. Sec.~\ref{sec:implication} extends the analysis made in previous section to the implications including the upcoming second-generation detectors KAGRA \cite{Somiya:2011np} and LIGO India \cite{Unnikrishnan:2013qwa}, taking account of the uncertainty of the coupling with magnetic fields. Finally, Sec.~\ref{sec:conclusion} is devoted to summary of important findings and conclusion. 
%
%
\section{Cross correlation analysis and impact of correlated noises}
\label{sec:formulation}



In this section, we begin by briefly reviewing the cross correlation analysis to detect the stochastic gravitational waves, and discuss the impact of nonvanishing correlated noise on the standard cross correlation analysis.

Let us denote the time-series data of the signal strain measured at $i$-th detector by $s_{i}(t)$, which is usually described by a sum of the gravitational-wave amplitude $h_{i}$ and the detector noise $n_i$: 
\begin{align}
 s_{i}(t)=h_{i}(t)+n_{i}(t). 
\end{align}
The stochastic gravitational waves (GWs) have a random nature, and only with single detector, it is generally difficult to distinguish the GW signal from instrumental noise. This is particularly the case when the amplitude of GWs is comparable or much smaller than that of the noise. In such a situation, one way to separate the stochastic signal from others is to use multiple set of output data obtained from the different detectors, and to take the cross correlation between them. If the origin of noises is associated with the local instrumental setup and there is no environmental correlation, we do not expect the sizable noise correlation between different detectors. In this case, the cross correlation is powerful and enables us to pick up the stochastic GW signals. 

Assuming that both the stochastic GW and noise obey stationary Gaussian random process, we define the cross-correlation statistic:
\begin{align}
 S=\int_{-T/2}^{T/2}dt \int_{-T/2}^{T/2}dt'\, s_{1}(t)s_{2}(t') Q(t-t'),
\label{eq:correlation_signal}
\end{align}
where $T$ is the observation time of the order of $1\,{\rm year} \approx 3.0 \times 10^{7}$ seconds, and $Q(t-t')$ is the optimal filter function to enhance the detectability of the gravitational-wave signal (see below). The expectation value (i.e., ensemble average) of this statistic then leads to 
\begin{align}
 \langle S \rangle  =\langle S_{\rm G}\rangle,
\label{eq:cross_correlation}
\end{align}
with $\langle S_{\rm G}\rangle$ given by
\begin{align}
\langle S_{\rm G} \rangle \equiv \int_{-T/2}^{T/2}dt \int_{-T/2}^{T/2}dt'\, 
\langle h_{1}(t)h_{2}(t') \rangle \,Q(t-t') \,.
\label{Sgw}
\end{align}
This is expressed in terms of the quantities in the Fourier domain as (e.g., Ref.~\cite{Allen:1997ad})
\begin{align}
 \langle S_{\rm G} \rangle = \frac{3H_{0}^{2}}{10\pi^{2}}T
\int_{0}^{\infty}\,df\, f^{-3}\, \Omega_{\rm gw}(f)\, \gamma_{1 2}^{\rm G}(f) 
\,\widetilde{Q}(f)\,,
\label{eq:S_G_Fourier}
\end{align}
where $H_0=100\,h\,$km\,s$^{-1}$\,Mpc$^{-1}$ is the Hubble parameter, $\Omega_{\rm gw}$ is the dimensionless quantity defined by the gravitational-wave energy density stored in a logarithmic frequency interval around $f$ divided by the critical energy density, and $\gamma_{1 2}^{\rm G}$ represents the coherence of the gravitational strains between the two separated detectors, referred to as the overlap reduction function (see Sec.~III-B of Ref.~\cite{Allen:1997ad} and Appendix B of Ref.~\cite{Flanagan:1993ix} for derivation). The function $\widetilde{Q}$ is the Fourier transform of the optimal filter function, and in deriving the expression above, we have assumed that the support of the filter function is sufficiently small in time domain compared to the observation time.

In contrast to the mean value $\langle S\rangle$ given above, the dispersion of the cross correlation statistic $S$, defined by $\sigma^{2} \equiv \langle S^{2} \rangle-\langle S \rangle^{2}$, would be dominated by the detector's noise. This is especially the case in the weak-signal limit ($h_i\ll n_i$). The signal-to-noise ratio for detecting the stochastic GWs is thus defined as 
\begin{align}
 {\rm SNR_{G}} \equiv \frac{\langle S_{\rm G} \rangle}{\sigma},
\label{eq:SNR_G}
\end{align}
where the explicit expression of $\sigma$ is given in the weak-signal limit 
 as follows:
\begin{align}
\sigma^{2} \simeq \frac{T}{2}
\int_{0}^{\infty}\,df\,P_{1}(f)\,P_{2}(f)\,|\widetilde{Q}(f)|^{2},
\label{eq:variance}
\end{align}
with the function $P_i$ being the instrumental noise spectrum for $i$-th 
detector. The filter function is still unknown at this moment, but it can be chosen so as to maximize this signal-to-noise ratio. For our interest in the weak-signal limit, the optimal filter is taken in the following functional form \cite{Allen:1997ad}: 
\begin{align}
 \widetilde{Q}(f)\propto\frac{\gamma_{1 2}^{\rm G}(f)\, \Omega_{\rm gw}(f)}{f^{3}\,P_{1}(f)\,P_{2}(f)}\,.
\label{eq:filter_function}
\end{align}

The discussions given above are the standard cases in the absence of noise correlation. For the ground-based detectors, the validity of this assumption is questionable, and the significance of the correlated noise has been pointed out (e.g., Refs.~\cite{Christensen:1992wi, Allen:1997ad}). The global magnetic field in the cavity formed by the surface of the Earth and ionosphere is known to be an important candidate, and it can cause the correlation between widely separated detectors by inducing forces on magnets mounted on the mirror control system. In particular, the stationary component of the magnetic field possesses the resonant peak structure at the low frequencies, referred to as the Schumann resonance \cite{1952ZNatA...7..149S, 1952ZNatA...7..250S}, and without perfectly shielding the magnetic fields, the correlated noise can potentially mask the stochastic GW signals at low-frequency bands.

In what follows, we focus on the correlated noise produced by the Schumann resonances, and present the basis for a quantitative estimate of its impact on the detection of GWs. For this purpose, we construct an analytic model of correlated noise. Let us divide the strain amplitude of the noise $n_{i}$ into two pieces: 
\begin{align}
 n_{i}(t)=n_{i}^{\rm I}(t)+n_{i}^{\rm B}(t).
\end{align}
Here, $n_{i}^{\rm I}(t)$ is the instrumental noise produced by local disturbances, and $n_{i}^{\rm B}(t)$ is the correlated noise induced by the global magnetic fields on the Earth. The second term produces a nonvanishing contribution to the cross correlation statistic, and with this term the expectation value now becomes $\langle S\rangle=\langle S_{\rm G}\rangle+\langle S_{\rm B}\rangle$ with $\langle S_{\rm B}\rangle$ given by
\begin{align}
 \langle S _{\rm B} \rangle = 
\int_{-T/2}^{T/2}dt\int_{-T/2}^{T/2}dt'\, \langle n_{1}^{\rm B}(t)n_{2}^{\rm B}(t') \rangle \,Q(t-t'). 
\label{eq:sb}
\end{align}
Unless the size of this term is quantitatively determined or predicted, we are unable to separately measure the GW signal. Since the correlated noise here is produced by a weak coupling of the magnetic field with mirror control system, one  can assume that $n_i^{\rm B}$ is linearly proportional to the magnetic field $B^a$ at $i$-th detector's position, ${\bm x}_i$. Then, in Fourier domain, the (stationary) correlated noise is generally expressed as
\begin{align}
\widetilde{n}_{i}^{\rm B}(f)=r_{i}(f)\,\left[{\widehat{\bm X}}_{i}\,\cdot \widetilde{{\bm B}}(f,{\bm x}_{i})\right].
\label{eq:conv_noise}
\end{align}
Here, the frequency-dependent quantity $r_i$ is called the transfer function, 
which characterizes the strength of the coupling between the detector and 
magnetic field, and the unit vector $\widehat{\bm X}_i$ describes the 
directional dependence of the coupling with magnetic field\footnote{In what follows, a hatted quantity implies a unit vector.}. Note in general that 
$\widehat{\bm X}_i$ is also given as a function of the frequency. These two quantities are basically determined by the details of detector setup especially around the magnets. Though it would be difficult to precisely determine their functional form from the first principle calculation, we can in principle calibrate them with the equipped sensors (magnetometers) that monitors the local environments. On the other hand, the magnetic fields $\widetilde{B}^{a}$ associated with Schumann resonances have a global coherence with the scale comparable to the Earth size, and the actual size of their correlation cannot be determined by a local experiment.

Adopting Eq.~(\ref{eq:conv_noise}), the expectation value $\langle S _{\rm B} \rangle$ is rewritten in terms of the Fourier-space quantities with
\begin{align}
\langle S _{\rm B} \rangle 
&= T\, \int_{0}^{\infty}\,df \,\,{\rm Re}\bigl[r_{1}^{\ast}(f)\,r_{2}(f)\,M_{1 2}(f)\bigr]\,\widetilde{Q}(f),  
\label{eq:snrBB2}
\end{align}
where the function $M_{12}$ is the correlated magnetic noise spectrum for a pair of detector, given by 
\begin{align}
M_{12}(f)&= \widehat{X}_{1,a}\widehat{X}_{2,b} \,\,
\langle \widetilde{B}^{a *}(f,{\bm x}_{1}) \widetilde{B}^b(f',{\bm x}_{2})\rangle',
\label{eq:M_12}
\end{align}
where angle bracket denotes an ensemble average and a prime on a correlator indicates that the Dirac factor $\delta(f-f')$ is dropped. The labels $a,\,b$ run from $1$ to $3$. Note that in deriving Eq.~(\ref{eq:snrBB2}), we assumed that the support of the integrand is well within the one determined by the observation time.

Eq.~(\ref{eq:M_12}) is very close to what has been measured by Ref.~\cite{2013PhRvD..87l3009T}. To be precise, Ref.~\cite{2013PhRvD..87l3009T} measured the magnetic field correlation between LIGO Hanford, Livingston, and Virgo detectors using the magnetometers inside the observatory buildings. While this is not a direct observation obtained from the interferometric signals, the measured data clearly show the Schumann resonance peaks. One important remark is that the amplitude and structure of resonance peaks vary with detector pair. While this could be partly ascribed to the seasonal variation, the nonuniform structure of the magnetic field can give a major contribution to the variation of Schumann resonance peaks. If this is the case, the geometrical configuration of each detector pair may play an important role to mitigate the impact of correlated noise. We will investigate this issue based on a simple analytic model.

\section{Analytic model of correlated noise}
\label{sec:analytic_model}

In this section, we present a simple analytic model of noise correlation, and derive the analytic expression of Eq.~(\ref{eq:snrBB2}) [see Eqs.~(\ref{eq:SB_discrete})-(\ref{eq:gamma_integral})]. We then study the properties of the correlated noise, and discuss the geometrical dependence of the detector pair. 

\subsection{Model description}
\label{property}

First recall that the Schumann resonances are represented by the random superposition of the electromagnetic waves propagating in the Earth-ionosphere cavity. These waves are steadily produced by excitation sources, and can be represented by the normal modes of the Earth-ionosphere waveguide. Primary natural source of Schumann resonances is the lightning discharges which frequently happen over the sky. Here, to simplify the analysis, we consider an idealistic model of resonant cavity made of two perfectly conducting, concentric spheres with radii $r=R_\oplus$ and $R_\oplus+a$, where $R_\oplus \approx 6,400$km is the radius of the Earth and $a\sim100$\,km is the height of the ionosphere from the surface of the Earth. According to the theory of waveguides \cite{1998clel.book.....J}, the electromagnetic waves are generally classified as two modes called transverse electric (TE) and transverse magnetic (TM) modes, for which the radial electric and magnetic field components vanish, respectively. Among these, the lowest frequencies of the TE modes become of the order of $c/a \sim 10^{3}$\,Hz, with the quantities $c$ being the light velocity. Hence, only the TM modes have relevant frequencies of the order of $c/R_\oplus \sim 10$\,Hz, at which the ground-based laser interferometers are most sensitive to the GW signals.

\begin{figure}[tb]

\vspace*{-1cm}
  
\begin{center}
\includegraphics[width=9cm,angle=0,clip]{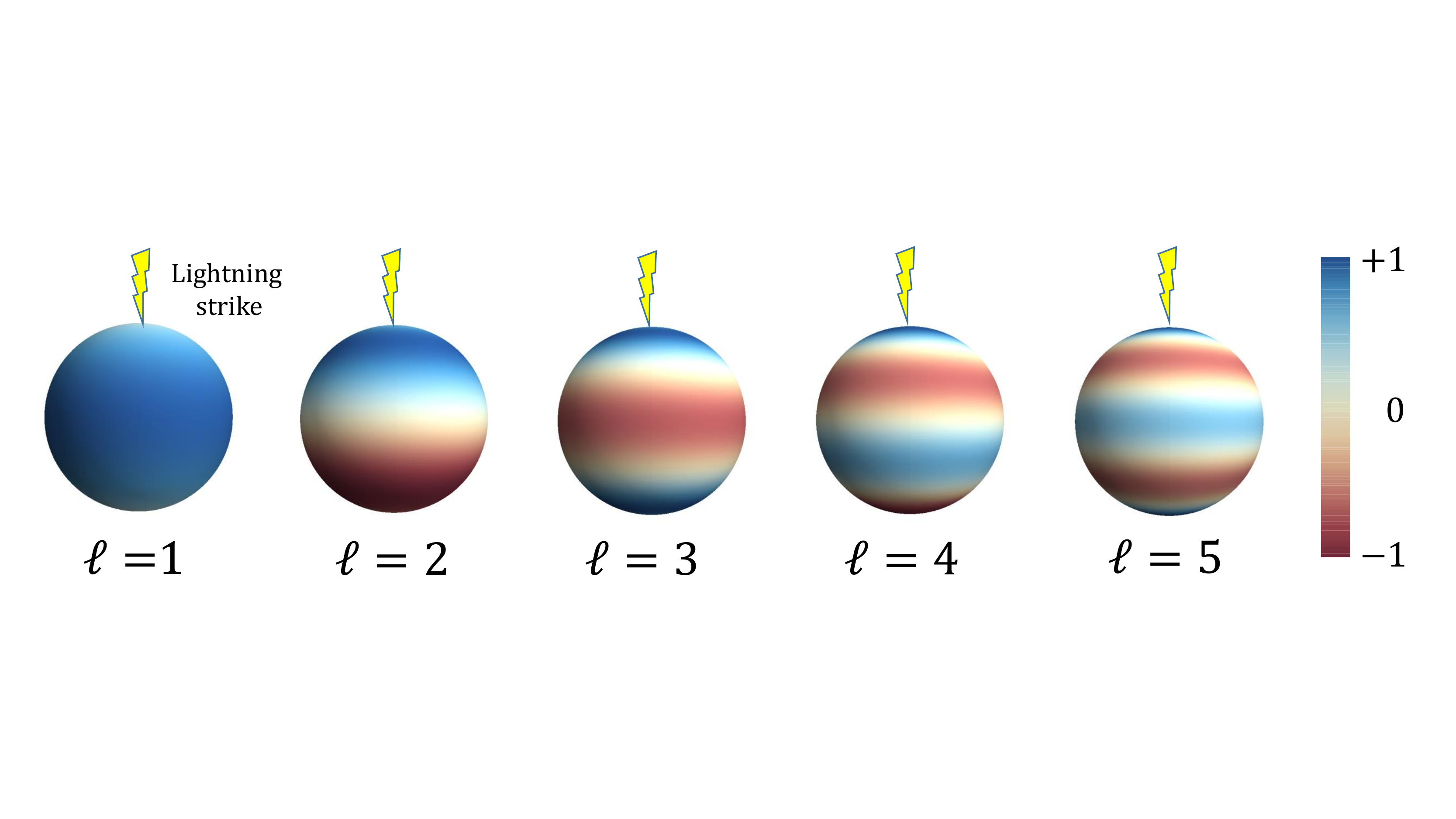}
\end{center}

\vspace*{-1.5cm}

\caption{Axisymmetric TM modes for the first five Schumann resonances, $\ell=1 \,\, {\rm to}\,\, 5$. The shade of color in each spherical plot indicates the behavior of $\mathcal{P}_{\ell}^{1}$ divided by each-$\ell$ maximum value. Blue and dark red in right indicator represents positive and negative value for $\mathcal{P}_{\ell}^{1}$, respectively. The darkest blue (:+1 in the indicator) is equivalent to maximum value.
}
\label{fig:single_mode}
\end{figure}

Let us then consider the property of a single TM mode. For a spherical geometry, the produced TM mode inside the cavity can be axisymmetric with respect to the lightning source. In Ref.~\cite{1998clel.book.....J}, the analytic form of the axisymmetric TM mode is presented (see Chap.8.9 of this textbook). Taking the center of the Earth as the origin of all position vectors and denoting the direction of the lightning source by $\widehat{\bm \Omega}$, the axisymmetric TM mode at given position ${\bm r}$ has nonvanishing component along the direction $\widehat{\bm \Omega}\times\widehat{\bm r}$. Explicit functional form of the TM mode is given by (see Fig.~\ref{fig:single_mode} for illustration) 
\begin{align}
& {\bm B}^{\rm TM}(t,\,{\bm r};\,\widehat{\bm \Omega})=B^{\rm TM}(t,\,{\bm r};\,\widehat{\bm \Omega})\,\,\widehat{\bm e}(\widehat{\bm \Omega}); 
\label{eq:mode_func}
\\
&\qquad\quad 
B^{\rm TM}=\,
\frac{u_{\ell}(r)}{r}\, \mathcal{P}_{\ell}^{1}(\widehat{\bm \Omega} \cdot \widehat{\bm r})\,
e^{-i \,2\pi f\, t},
\nonumber\\
&\qquad\quad 
\widehat{\bm e}=\frac{\widehat{\bm \Omega}\times\widehat{\bm r}}{|\widehat{\bm 
 \Omega}\times\widehat{\bm r}|},
\nonumber
\end{align}
where the function $\mathcal{P}_{\ell}^{1}$ is the associated Legendre polynomials\footnote{The associated Legendre polynomials used in this paper, $\mathcal{P}_\ell^m$, are related to the Legendre polynomials, $\mathcal{P}_\ell$, through
\begin{align}
  \mathcal{P}_\ell^m(\mu)=(1-\mu^2)^{m/2}\,\frac{d^m \mathcal{P}_\ell(\mu)}{d\mu^m}.
\end{align}
There is another definition frequently used in the literature, for which the factor $(-1)^m$ is multiplied in the above relation. If the latter definition is adopted, the sign of the terms involving the function $\mathcal{P}_\ell^1$ has to be flipped for the analytic expressions presented below. },  
 and $u_{\ell}(r)$ is the radial-mode function which satisfies 
\begin{align}
& \left[\frac{d^{2}}{dr^{2}}+\{\omega_\ell(r)\}^2\right]\,u_{\ell}(r)=0\,;
\nonumber\\
&\qquad\qquad 
\omega_\ell(r)=\sqrt{\frac{(2 \pi f)^{2}}{c^{2}}
-\frac{\ell(\ell+1)}{r^{2}}},
\end{align}
with the multipole $\ell=1,2,\cdots$ characterizing the angular dependence of the modes. The characteristic frequencies of the Schumann resonances are determined from the solution of this equation by imposing the boundary condition from the perfectly conducting walls, $du_\ell/dr=0$ at $r=R_\oplus,\,\,R_\oplus +a$. While it generally requires us to solve the transcendental equation, in the case of $a/R_\oplus\ll1$, the solution satisfying the boundary condition is approximately described by $u_\ell\simeq A\cos\,[\omega_\ell(R_\oplus)\,(r-R_\oplus)]$ with $\omega_\ell(R_\oplus)=m\,\pi/a$ and $m=0,1,2,\cdots$. Among various modes with different integer $m$, relevant low-frequency modes arise only from $m=0$, which leads to 
\begin{align}
f_{\ell} \simeq \frac{c}{2 \pi R_{\oplus}} \sqrt{\ell(\ell+1)}. 
\label{eq:Schumann_fr}
\end{align}
With this relation, the first three resonant frequencies are predicted to be $f_\ell=10.6,\,\,18.3,\,\,25.9$ Hz, which are known to slightly differ from the actual measured frequencies. We will later discuss this issue.

Using the single TM mode given above, we now express the global magnetic field for the Schumann resonances, which is given as the random superposition of TM modes produced by the widely distributed excitation sources. Summing up the contributions over the sky, the magnetic field at the $i$th detector position, ${\bm r}_i=R_\oplus\,\widehat{\bm r}_i$, becomes
\begin{align}
{\bm B}^{\rm SR}(t,\,{\bm r}_{i}) &= \sum_{\ell=1}^{\infty}\, \int_{S^2} d^{2}\widehat{\bm \Omega}\,\,{\bm B}^{\rm TM}(t,\,R_\oplus\widehat{\bm r}_i;\,\widehat{\bm \Omega})\,
 + {\rm c.c.} ,
\label{eq:mode_expansion}
\end{align}
where the mode ${\bm B}^{\rm TM}$ depends implicitly on the multipole $\ell$, and we sum up all possible multipoles. The above expression can be recast in terms of the quantities in Fourier domain as
\begin{align}
&{\bm B}^{\rm SR}(t,\,{\bm r}_{i}) = \sum_{\ell=1}^{\infty}\, \Delta f \,\,
\widetilde{\bm B}^{\rm SR}(f_\ell,{\bm r}_i)\,e^{-i\,2\pi\,f_\ell t} + {\rm c.c.}\,;
\nonumber\\
&\qquad 
\widetilde{\bm B}^{\rm SR}(f_\ell,\,{\bm r}_i) = 
\sqrt{\frac{(2\ell+1)(\ell-1)!}{4\pi\,(\ell+1)!}}
\nonumber\\
&\qquad\quad\times
\,\int_{S^2} d^{2}\widehat{\bm \Omega}\,\,
\widetilde{B}(f_\ell, \widehat{\bm \Omega})\, \mathcal{P}_\ell^1 (\widehat{\bm \Omega} \cdot \widehat{\bm r}_i)\,\widehat{\bm e}_i(\widehat{\bm \Omega}).
\label{eq:SR_modes}
\end{align}
The quantity $\widehat{\bm e}_i$ is the unit vector given by 
\begin{align}
&\widehat{\bm e}_i(\widehat{\bm \Omega})=\frac{\widehat{\bm \Omega}\times\widehat{\bm r}_i}{|\widehat{\bm \Omega}\times\widehat{\bm r}_i|}.
\label{eq:unit_vect_ei}
\end{align}
Here, the quantity $\widetilde{B}(f_\ell,\widehat{\bm \Omega})$ represents the 
amplitude of the TM mode associated with a lightning source at $\widehat{\bm 
\Omega}$. In the above, the dimensional fudge factor $\Delta f$ is introduced so 
that $\widetilde{\bm B}^{\rm SR}$ or $\widetilde{B}$ is defined similarly to what is normally seen in the continuous limit. 
Since Eq.~(\ref{eq:SR_modes}) is evaluated at $r=R_\oplus$, the dependence of the radial-mode function is now meaningless, and is absorbed into the amplitude of TM mode with the appropriate normalization factor.

In Eq.~(\ref{eq:SR_modes}), the stochastic nature of the magnetic field is encapsulated in the amplitude of each TM mode, $\widetilde{B}$. The statistical property of the random amplitude $\widetilde{B}$ is characterized by the power spectral density $P_{\rm B}$. While precise functional form of it would be, in reality, complicated according to the geographical and climate reasons, we here, for simplicity, assume that the amplitude $\widetilde{B}$ is statistically isotropic. Then, the (single-sided) power spectral density is defined by
\begin{align}
 \langle {\widetilde B}^{*}(f_\ell,\widehat{\bm \Omega})\,{\widetilde B}(f_{\ell '}, \widehat{\bm \Omega}')\rangle
= \frac{\delta^{2}(\widehat{\bm \Omega},\widehat{\bm \Omega}')}{4\pi} \,\frac{\delta_{\ell \ell'}}{\Delta f}\,\frac{P_{\rm B}(f_{\ell})}{2}, 
\label{eq:expectation_valueB}
\end{align}
where we also introduced $\Delta f$ to recover the correct physical dimension in the continuous limit. The assumptions of isotropy and stationary random process is perhaps rather idealistic, and the extension to a realistic case may have to be considered in more quantitative study. Nevertheless, we will see later that even the simple description of magnetic field can capture several important properties, and accounts for what have been found by Refs.~\cite{2013PhRvD..87l3009T,2014PhRvD..90b3013T}.

\begin{figure}[tb]
\includegraphics[width=8.5cm,angle=0,clip]{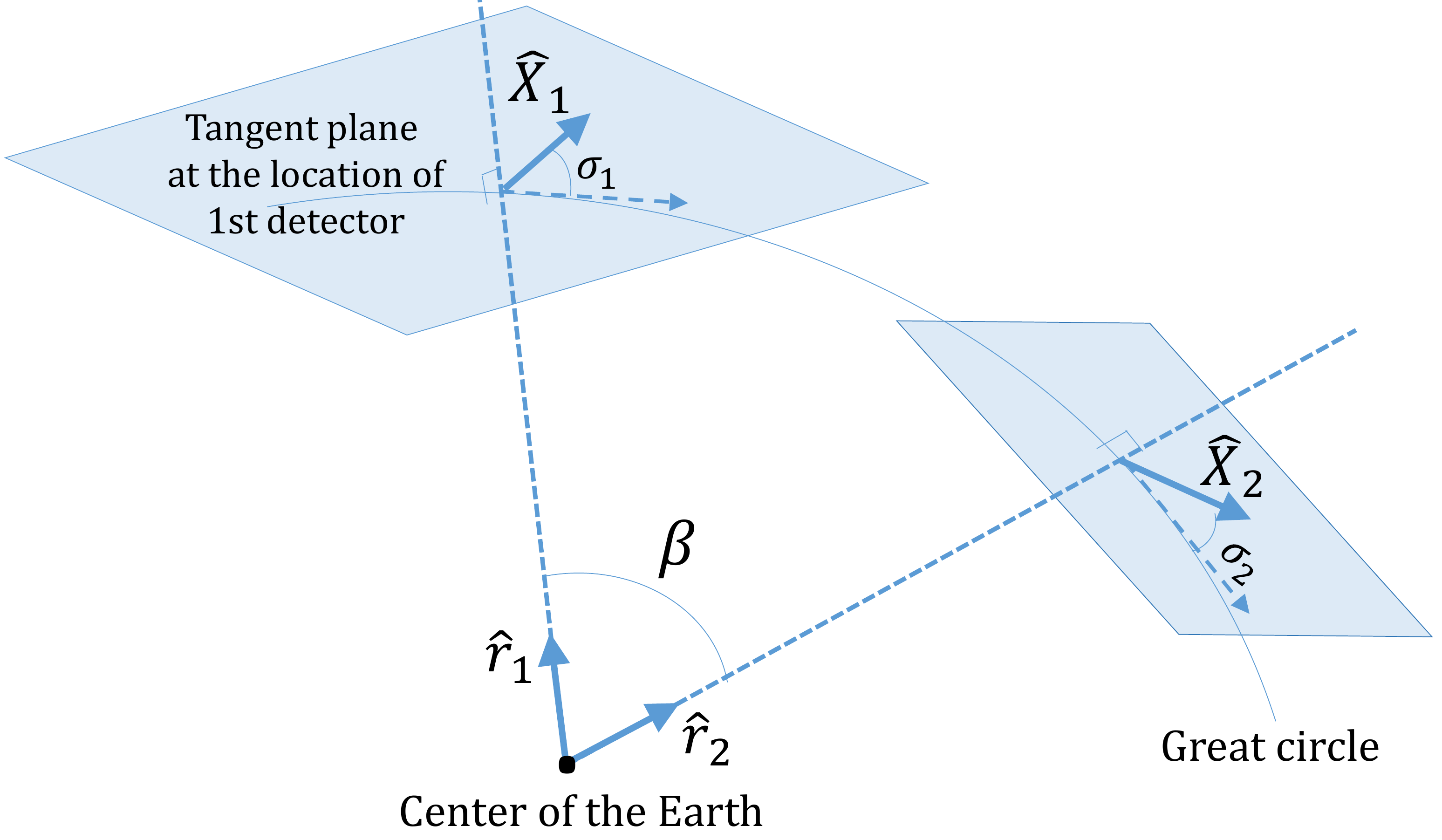}
\caption{Description of geometrical configuration for two detectors on the great circle. $\widehat{\bm X}_{i}$ is the projection vector embedded in a tangent plane at the location of $i$-th detector. The direction of $\widehat{\bm X}_{i}$ is characterized by the misalignment angle $\sigma_{i}$ with respect to a great circle connecting two detectors. The relative distance between 1st and 2nd detectors is characterized by $\beta$ which corresponds to $\cos^{-1} (\widehat{\bm r}_{1} \cdot \widehat{\bm r}_{2})$.}
\label{fig:great_circle}
\end{figure}

Adopting the model of global magnetic field presented above,  
we now derive the expression of correlated noise, $\langle S_{\rm B}\rangle$ in Eq.~(\ref{eq:sb}). Using Eqs.~(\ref{eq:conv_noise}) and (\ref{eq:expectation_valueB}), substituting Eq.~(\ref{eq:SR_modes}) into the definition of $\langle S_{\rm B}\rangle$ leads to the discrete version of Eq.~(\ref{eq:snrBB2}): 
\begin{align}
& \langle S_{\rm B} \rangle 
= T
\sum_{\ell=1}^{\infty} \,\Delta f\,\,{\rm Re}\bigl[r_1^{*}(f_{\ell})\,r_2(f_{\ell})\bigr]
\,M_{12}(f_\ell)\,\widetilde{Q}(f_\ell)\,,
\label{eq:SB_discrete}
\end{align}
with the explicit expression for the correlated magnetic noise spectrum $M_{12}$ given by
\begin{align}
M_{12}(f_\ell)=\frac{1}{8\pi}\,P_{\rm B}(f_\ell)\,\gamma_{\ell}^{\rm B}(\widehat{\bm r}_{1},\widehat{\bm r}_{2}). 
\label{eq:M12_model}
\end{align}
Here, we defined the function $\gamma_\ell^{\rm B}$, which characterizes the coherence of the global magnetic field: 
\begin{align}
& \gamma_{\ell}^{\rm B}(\widehat{\bm r}_{1},\widehat{\bm r}_{2}) = 
\frac{(2\ell+1)}{2\pi}\frac{(\ell-1)!}{(\ell+1)!} \nonumber \\
\times &
\int_{S^{2}}\,d^2{\widehat {\bm \Omega}} 
\,\mathcal{P}_{\ell}^{1}(\widehat{\bm \Omega} \cdot \widehat{\bm r}_{1})
\,\mathcal{P}_{\ell}^{1}(\widehat{\bm \Omega} \cdot \widehat{\bm r}_{2}) 
\, \{{\widehat{\bm e}}_{1}(\widehat{\bm \Omega}) \cdot \widehat{\bm X}_{1}\}\,
\, \{{\widehat{\bm e}}_{2}(\widehat{\bm \Omega}) \cdot \widehat{\bm X}_{2}\}\,.
\label{eq:gamma_integral}
\end{align} 
Denoting the directional cosine $(\widehat{\bm r}_1\cdot\widehat{\bm r}_2)$ by $\mu$, the integral in Eq.~(\ref{eq:gamma_integral}) is performed analytically, to obtain
\begin{align}
  &\gamma_{\ell}^{\rm B}(\widehat{\bm r}_{1},\widehat{\bm r}_{2}) =
  \frac{2\ell+1}{2\pi}\frac{(\ell-1)!}{(\ell+1)!} \nonumber\\
& \times \Bigl[
F_\ell(\mu)\,\left\{\mu\,(\widehat{\bm X}_1 \cdot \widehat{\bm X}_2)
-(\widehat{\bm r}_{2} \cdot \widehat{\bm X}_1) \, (\widehat{\bm r}_{1} \cdot \widehat{\bm X}_2) \right\}
\nonumber \\
&
-G_\ell(\mu)\,\left\{(\widehat{\bm r}_1 \times \widehat{\bm r}_2) \cdot \widehat{\bm X}_1 \right\}
\, \left\{(\widehat{\bm r}_1 \times \widehat{\bm r}_2) \cdot \widehat{\bm X}_2 \right\}
\Bigr],
\label{eq:Gamma}
\end{align}
with the functions $F_\ell$ and $G_\ell$ given by
\begin{align}
&F_\ell(\mu) = -\frac{4\pi}{2\ell+1}\left[(\ell+1)\mathcal{P}_{\ell+1}(\mu)-\frac{\mu}{\sqrt{1-\mu^{2}}}\mathcal{P}_{\ell+1}^{1}(\mu)\right],
\label{eq:fl}
\\
&G_\ell(\mu) =  -\frac{4\pi}{2\ell+1}\,
\Biggl[(\ell+1)(\ell+2)\, \frac{\mu}{1-\mu^2}\, \mathcal{P}_{\ell+1}(\mu)
\nonumber \\
& \quad\quad\quad
+\frac{\ell-(\ell+2) \mu^{2}}{(1-\mu^{2})^{3/2}}\,\mathcal{P}_{\ell+1}^{1}(\mu)\Biggr]\,.
\label{eq:gl}
\end{align}
Derivation of Eq.~(\ref{eq:Gamma}) is presented in detail in Appendix \ref{analytic_gamma}.

The function $\gamma_\ell^{\rm B}$ is analogous to the overlap reduction function $\gamma_{12}^{\rm G}$ given in Eq.~(\ref{eq:S_G_Fourier}), and we see from Eq.~(\ref{eq:Gamma}) that its behavior is determined by the geometric configuration of the detector pair ($\widehat{\bm r}_i$) through the directional-dependent coupling with magnetic field, characterized by $\widehat{\bm X}_i$. Thus, the spectrum $M_{12}$ can vary not only with spectral property of the magnetic field but also with geometrical configuration of detector pair. 
We will investigate the behavior of the function $\gamma_\ell^{\rm B}$ in next subsection.

\begin{figure*}[!htb]
\begin{center}
\includegraphics[width=8.cm,angle=0,clip]{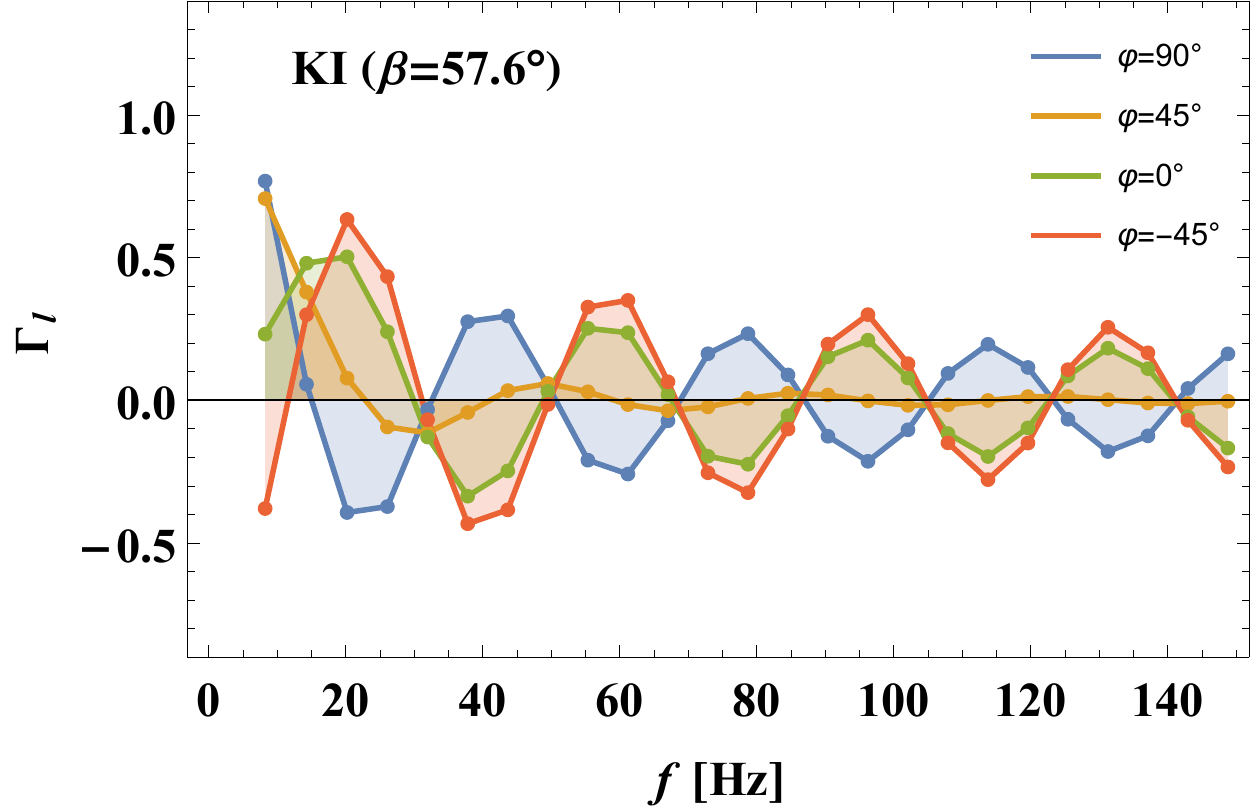}
\hspace{-15.02cm}
\includegraphics[width=8.cm,angle=0,clip]{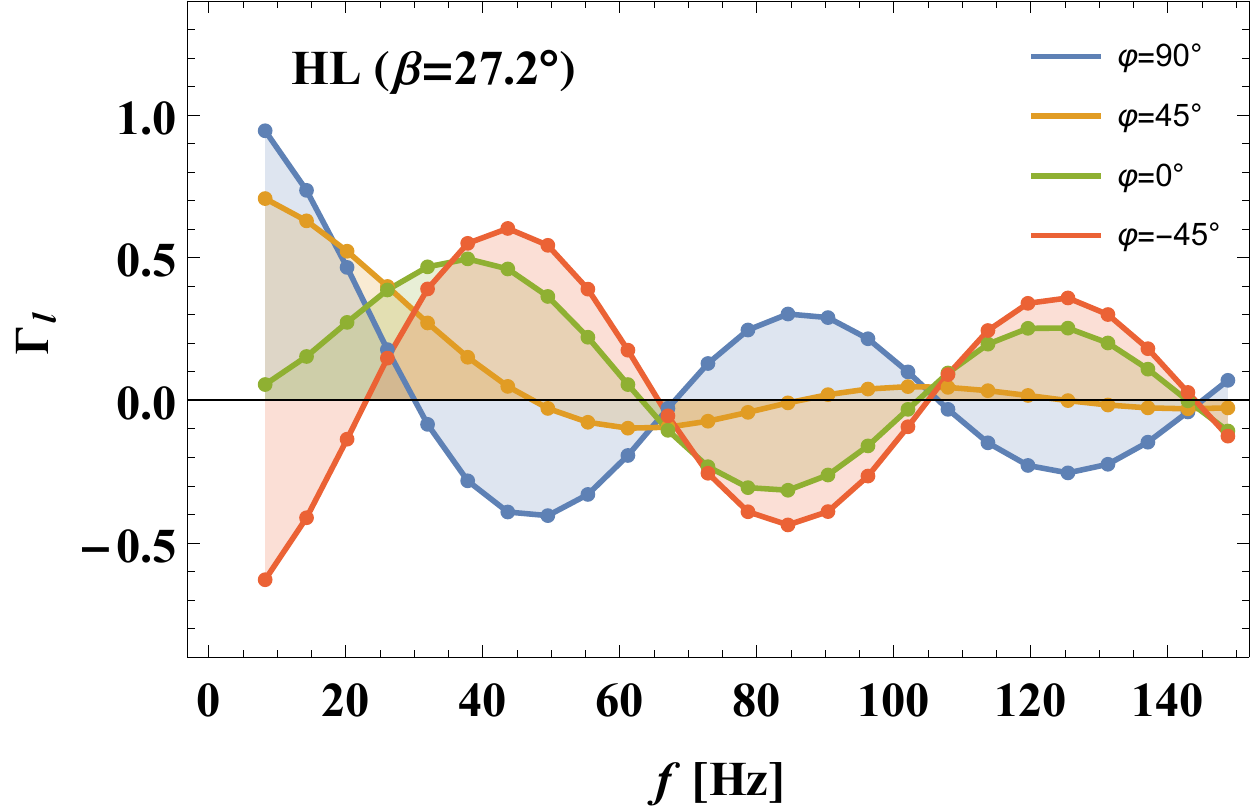}\\
\vspace{-1cm}
\includegraphics[width=8.cm,angle=0,clip]{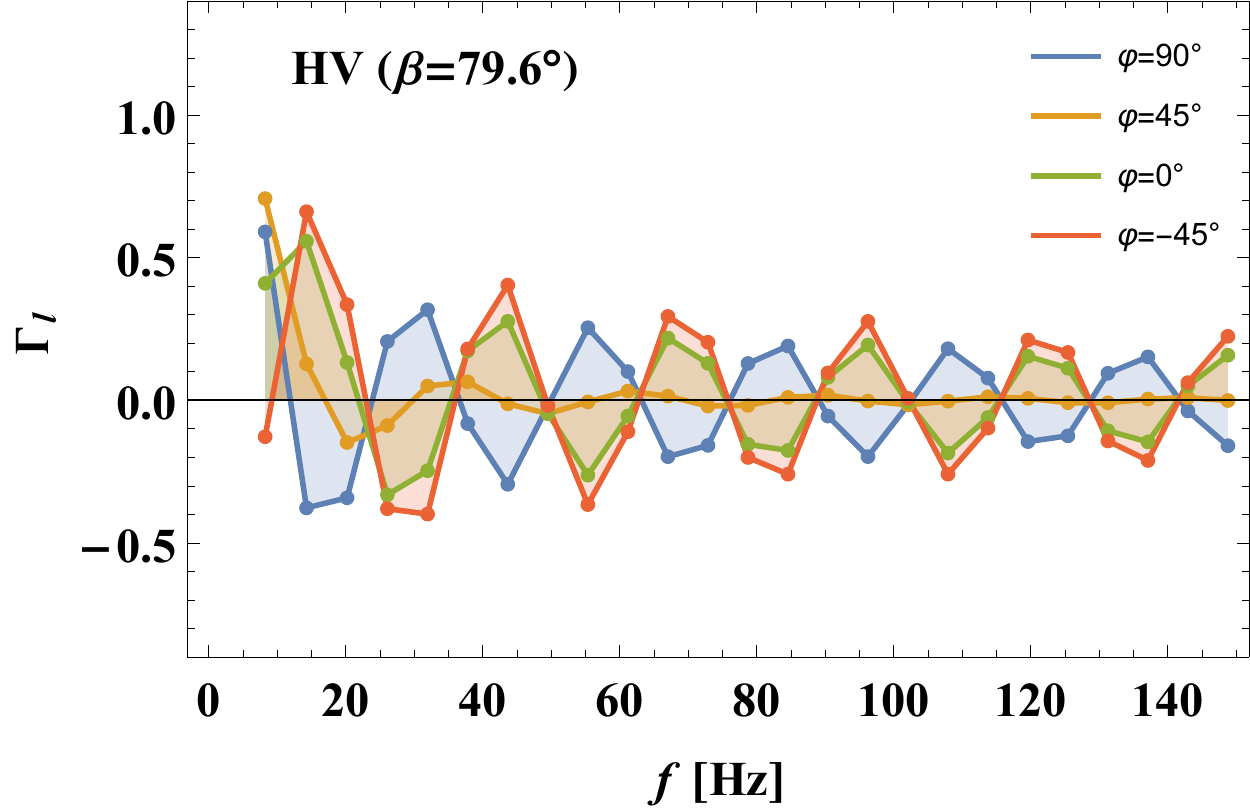}
\hspace{-15.02cm}
\includegraphics[width=8.cm,angle=0,clip]{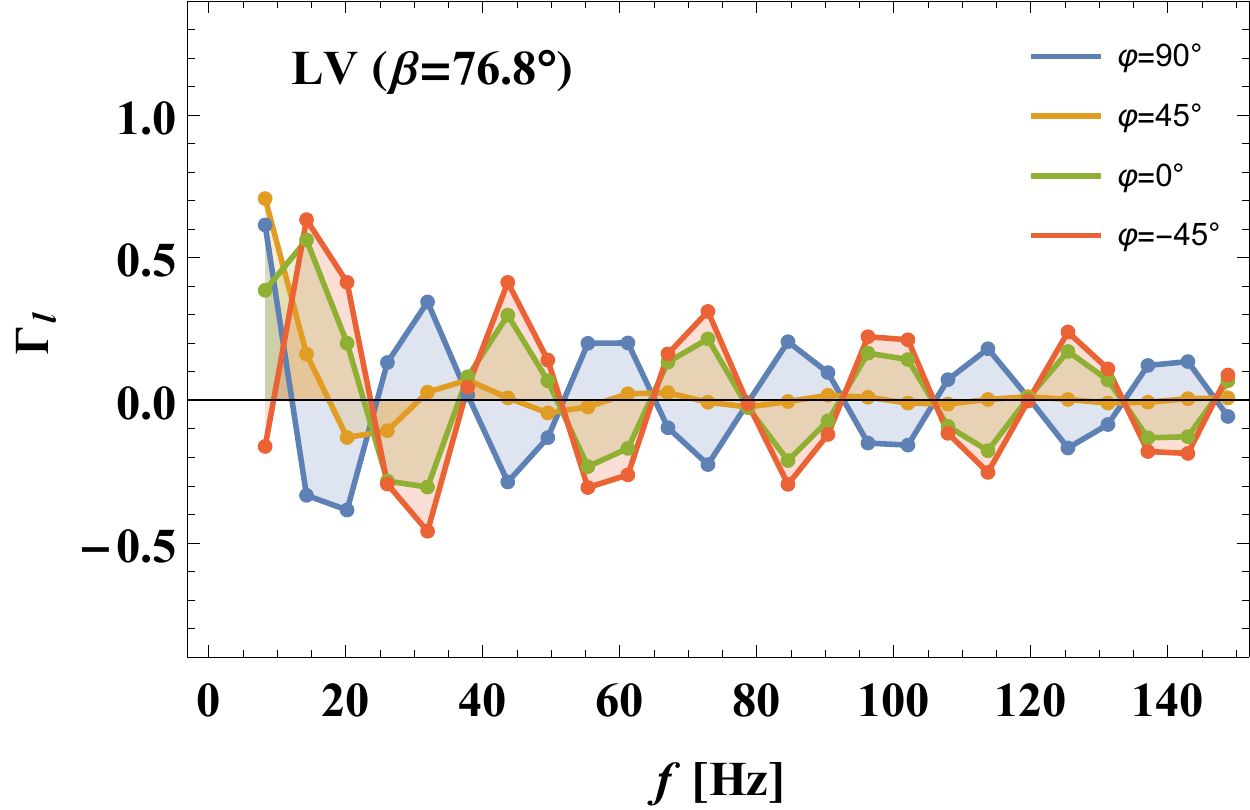}\\
\vspace{-1cm}
\includegraphics[width=8.cm,angle=0,clip]{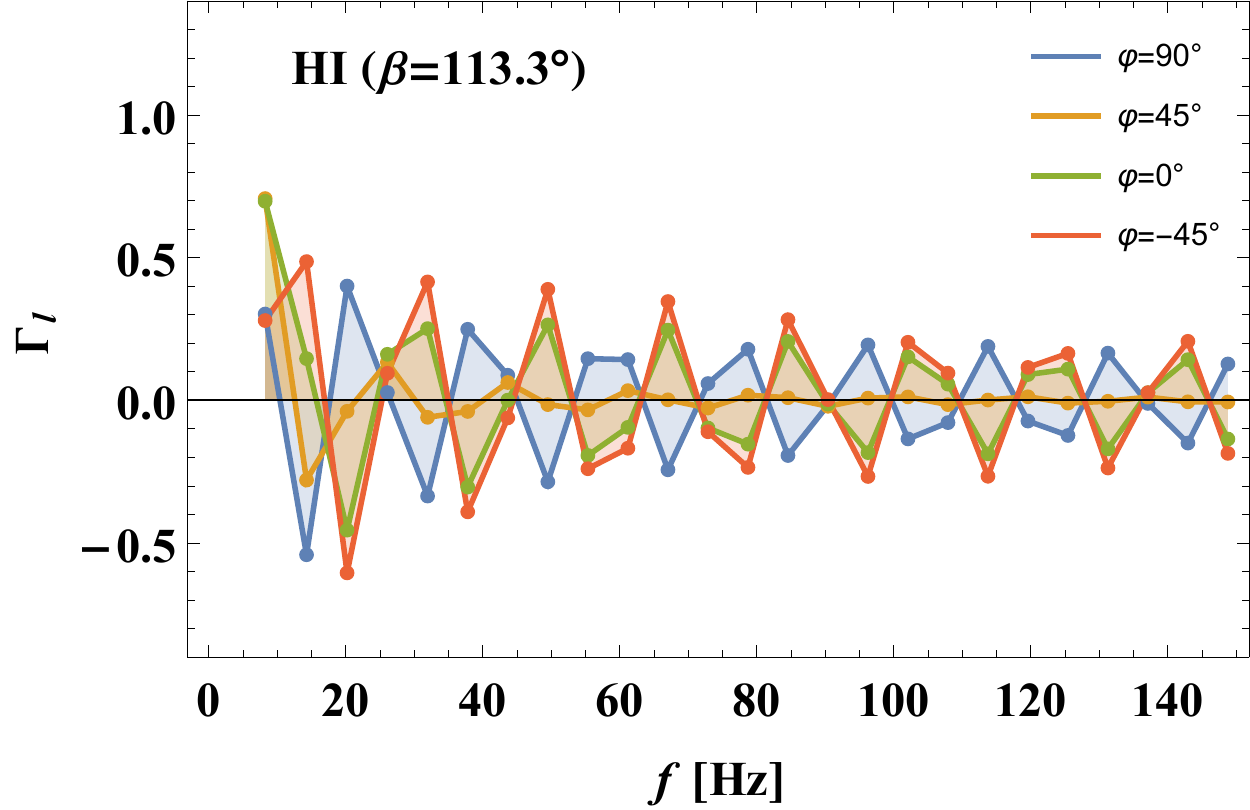}
\hspace{-15.02cm}
\includegraphics[width=8.cm,angle=0,clip]{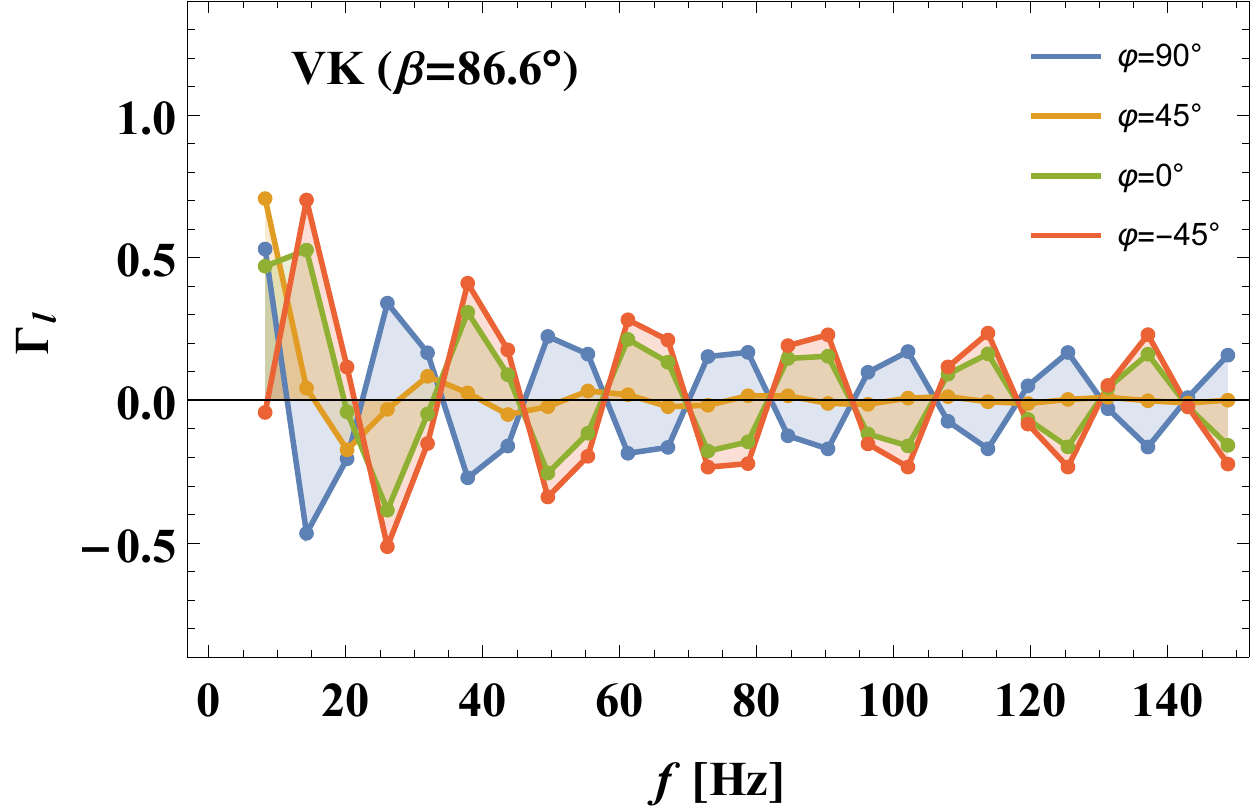}
\end{center}
\caption{Reduced amplitude of $\gamma_{\ell}^{\rm B}$ as a function of the rescaled frequency $0.78 f_{\ell}$ for LIGO Hanford-Livingston (HL, top-left), KAGRA-LIGO India (KI, top-right), LIGO Livingston-Virgo (LV, middle-left), LIGO Hanford-Virgo (HV, middle-right), Virgo-KAGRA (VK, bottom-left) and LIGO Hanford-India (HI, bottom-right). The phase angle $\varphi$ given by Eq.(\ref{eq:phase_angle}) is related to the relative orientation with respect to two projection vectors. As the separation angle $\beta$ increases (from top-left panel to bottom-right panel), $\Gamma_{\ell}$ rapidly oscillates over a wide range of the frequencies. 
}
\label{fig:normalized_gamma}
\end{figure*}

\subsection{Properties of correlated noise}
\label{property}

In this subsection, we elucidate the basic properties of the noise correlation in the simple analytic model. As we see in previous subsection, the key quantity is the function $\gamma_\ell^{\rm B}$ given at Eq.~(\ref{eq:Gamma}), which appears in the magnetic noise spectrum, $M_{12}$ [see Eq.~(\ref{eq:M12_model})]. Given the projection vector $\widehat{\bm X}_i$ at each detector and the directional cosine between the detector pair, the functional form of $\gamma_\ell^{\rm B}$ is uniquely specified, and is given as function of $f_\ell$. To see the dependence of these quantities clearly, we may rewrite the expression of $\gamma_\ell^{\rm B}$ in more compact form. 

Consider the great circle connecting the detector pair, as shown in Fig.~\ref{fig:great_circle}\footnote{This is basically the same characterization as first used in the overlap reduction function, $\gamma_{12}^{\rm G}$ (e.g., Refs.~\cite{Flanagan:1993ix,Seto:2008sr}).}. 
 Viewing this great circle from the center of the Earth, the size of the great circle is characterized by the angle $\beta$, which corresponds to $\cos^{-1}(\widehat{\bm r}_1\cdot\widehat{\bm r}_2)$. We then define the misalignment angle of the projection vector $\widehat{\bm X}_i$ with respect to the great circle, which we denote by $\sigma_i$. With these definitions, the vectors $\widehat{\bm X}_i$, which lies at the tangent plane on the sphere at ${\bm r}_i$, are expressed as
\begin{align}
  &\widehat{\bm X}_{1}=\cos (\sigma_{1})\,
\left(
  \begin{array}{c}
    \sin(\beta/2)
\\
0
\\
-\cos (\beta/2)
  \end{array}
  \right)+
  \sin (\sigma_{1}) \,\left(
\begin{array}{c}
  0
  \\
  1
  \\
  0
\end{array}
\right),
\label{eq:align1}
\\
 &\widehat{\bm X}_{2}=\cos (\sigma_{2})\,
 \left(
 \begin{array}{c}
   -\sin(\beta/2)
   \\
   0
   \\
   -\cos(\beta/2)
 \end{array}
 \right)
   +
   \sin (\sigma_2)\,\left( 
 \begin{array}{c}
   0
   \\
   1
   \\
   0
 \end{array}
 \right).
\label{eq:align2}
\end{align}
Substituting  Eqs.~(\ref{eq:align1}) and (\ref{eq:align2}) into Eq.(\ref{eq:Gamma}), after some algebra, the expression of $\gamma_\ell^{\rm B}$ is greatly simplified, and we have
\begin{align}
\gamma_{\ell}^{\rm B}(\widehat{\bm r}_{1},\widehat{\bm r}_{2})
=A_\ell(\beta) \,\cos(2\delta_{\rm B}) +B_\ell(\beta)\,\cos(2\Delta_{\rm B}),
\label{eq:gamma_analytic}
\end{align}
with the angles $\delta_{\rm B}$ and $\Delta_{\rm B}$ defined by 
\begin{align}
 \delta_{\rm B} \equiv \frac{\sigma_1 - \sigma_2}{2}\,, \quad \Delta_{\rm B} \equiv \frac{\sigma_1 + \sigma_2}{2}.
\label{eq:two_angles}
\end{align}
Here, the functions $A_\ell$ and $B_\ell$ are given by 
\begin{align}
& A_\ell(\beta)=\frac{2\ell+1}{2\pi}\frac{(\ell-1)!}{(\ell+1)!}
\nonumber\\
&\quad\times
\left\{F_\ell(\cos\beta)\,\cos^2\left(\frac{\beta}{2}\right)
-\frac{1}{2}\,G_\ell(\cos\beta)\,\sin^2\beta\right\},
\label{eq:A_ell}\\
& B_\ell(\beta)=\frac{2\ell+1}{2\pi}\frac{(\ell-1)!}{(\ell+1)!}
\nonumber\\
&\quad\times
\left\{F_\ell(\cos\beta)\,\sin^2\left(\frac{\beta}{2}\right)
+\frac{1}{2}\,G_\ell(\cos\beta)\,\sin^2\beta\right\}.
\label{eq:B_ell}
\end{align}

The expression given above helps to sort out the geometric dependence of the detector pair from the local properties of the magnetic-field coupling. From Eq.~(\ref{eq:gamma_analytic}), notable features of the correlated noise can be found for several specific setup. 

\begin{itemize}
 \item {\bf Colocated detectors}:  Taking the limit $\beta\to0^\circ$, one obtains $A_\ell\to1$  and $B_\ell\to0$, which lead to
\begin{align}
  \gamma_{\ell}^{\rm B}(\widehat{\bm r}_{1},\widehat{\bm r}_{2})|_{\beta = 0^{\circ}} = \cos(2\delta_{\rm B}).
\label{eq:gamma_beta0}
\end{align}
Thus, independent of the multipole $\ell$ or frequency $f_\ell$ [see Eq.~(\ref{eq:Schumann_fr})], the colocated detectors can be largely affected by the correlated noise. If the projection vectors of the magnetic coupling are aligned to the great circle of the detector pair, we have $\gamma_\ell^{\rm B}=1$. 

\item {\bf Antipodal detectors}: Consider next a pair of detectors located at 
       antipodal position. Taking the limit $\beta\to 180^\circ$, it is shown 
       in Appendix \ref{appendix:af} that  $A_\ell$ vanishes and $B_\ell$ becomes $(-1)^{\ell+1}$. We thus obtain
\begin{align}
 \gamma_{\ell}^{\rm B}(\widehat{\bm r}_{1},\widehat{\bm r}_{2})|_{\beta =  180^{\circ}}=(-1)^{\ell+1}\cos(2\Delta_{\rm B}).
\label{eq:gamma_B_antipodal}
\end{align}
That is, even the furthest separated detectors can have a potential of a large correlated noise, and the impact may be non-negligible even at high-frequency range. This is rather contrasted to the overlap reduction function of stochastic GWs, $\gamma_{12}^{\rm G}$, whose amplitude decays rapidly as increasing the distance between the pair of detectors \cite{Flanagan:1993ix,Allen:1997ad,Seto:2008sr}. 
It directly reflects the fact that the Schumann resonances are the phenomena associated with the global magnetic fields on the Earth.

\item {\bf Nulling configuration}: In general cases with $0^\circ< \beta<180^\circ$, the functions $A_\ell$ and $B_\ell$ show oscillatory behavior with slight different phases over a wide frequency range, and thereby the detectors can be sensitively affected by the Schumann resonances. Still, however, there is a special configuration for which we can always set $\gamma_\ell^{\rm B}=0$, irrespective of the geometry of detector pair and frequency/multipoles. Such a condition is given by  
\begin{align}
 \cos(2\delta_{\rm B})=\cos(2\Delta_{\rm B})=0.
\label{eq:null_condition}
\end{align}
The configuration satisfying Eq.~(\ref{eq:null_condition}) can be realized when 
       the projection vector for the local magnetic coupling at each detector, 
       $\widehat{\bm X}_i$, points to the direction either parallel or normal to the great circle connecting the detector pair and the angles $\sigma_i$ satisfy $(\sigma_1-\sigma_2)= (m+1/2) \pi$ for integer $m$. 
This property comes from the statistical isotropy of the magnetic sources and axisymmetry of the TM modes. 
Although the projection vector is generally uncertain and is hard to measure, the existence of the nulling configuration may play an important role to mitigate the impact of noise correlation.  
\end{itemize}

Finally, to get an insight into generic behaviors of $\gamma_\ell^{\rm B}$, we 
consider the specific detector pairs for second-generation detectors, and 
examine the geometric dependence of correlated noise. 
Fig.~\ref{fig:normalized_gamma} presents the {\it reduced amplitude} of 
$\gamma_\ell^{\rm B}$ as function of frequency $f_\ell$. Based on the basic 
parameters listed in Table \ref{position_separation}, the results are 
particularly shown for LIGO Hanford-Livingston (HL, top-left), KAGRA-LIGO India 
(KI, top-right), LIGO Livingston-Virgo (LV, middle-left), LIGO Hanford-Virgo (HV, middle-right),
Virgo-KAGRA (VK, bottom-left), and LIGO Hanford-India (HI, bottom-right) pairs. Here, the reduced amplitude, $\Gamma_\ell$, is defined by 
\begin{align}
\Gamma_{\ell}(\beta,\varphi) \equiv 
 \frac{\gamma_{\ell}^{\rm B}(\widehat{\bm r}_{1},\widehat{\bm r}_{2})}
{ \sqrt{\cos^{2}(2 \delta_{\rm B})+\cos^{2}(2 \Delta_{\rm B})}},
\end{align}
with the phase angle $\varphi$ defined by 
\begin{align}
 \tan(\varphi)=\frac{\cos(2\delta_{\rm B})}{\cos(2\Delta_{\rm B})}.
\label{eq:phase_angle}
\end{align}
Apart from the overall amplitude, the frequency dependence of the function $\gamma_\ell^{\rm B}$ is wholly encapsulated in the function $\Gamma_\ell$. Since $\Gamma_\ell$ varies with the phase angle $\varphi$,  we examine in Fig.~\ref{fig:normalized_gamma} several cases with different values of $\varphi$. Note that in plotting the results, we adopt the rescaled frequency relation, $f_\ell'=0.78\,f_\ell$,  rather than the one given by Eq.~(\ref{eq:Schumann_fr}), which is empirically known to match the observed Schumann resonance frequencies \cite{1998clel.book.....J}.

Fig.~\ref{fig:normalized_gamma} reveals that the function $\Gamma_\ell$ or $\gamma_\ell^{\rm B}$ exhibits slowly oscillatory behavior for a close pair of detectors (i.e., HL), but as increasing the angle $\beta$ (or increasing separation of the detector pair), it becomes rapidly oscillating function over a wide range of the frequencies, approaching the limiting case in Eq.~(\ref{eq:gamma_B_antipodal}). One notable property may be that the oscillatory phases do not shift significantly with the phase angle $\varphi$, and thereby the zero-crossing points mostly remain the same. This implies that the oscillatory feature is a generic outcome of the global magnetic field, and it would appear in the actual noise correlation.


\section{Comparison to empirical estimates}
\label{sec:comparison}

The analytic noise model in Sec.~\ref{sec:analytic_model} is constructed based on several assumptions and simplification for global magnetic fields, which might be idealistic and inappropriate to investigate a more quantitative aspect of noise correlation. In this respect, the present model can be used only for a qualitative study of noise correlation. Nevertheless, the present model still captures several important properties, for which we can even access the quantitative estimates. To show the significance of this, in this section, we compare the analytic model predictions with the results obtained by Refs.~\cite{2013PhRvD..87l3009T, 2014PhRvD..90b3013T}. The quantities to be compared are the correlated magnetic noise spectrum, $M_{12}$, and the signal-to-noise ratio of the contamination by the correlated magnetic noise\footnote{This is the same quantities as used in Ref.~\cite{2014PhRvD..90b3013T}, although they adopt a different notation, ${\rm SNR}_{\rm M}$. }, 
\begin{align}
 {\rm SNR}_{\rm B}\equiv \frac{\langle S_{\rm B}\rangle}{\sigma}, 
\label{eq:def_SNR_B}
\end{align}
as defined similarly to the GW case [Eq.~(\ref{eq:SNR_G})]. The former has been first measured at LIGO Hanford/Livingston and Virgo detectors through the on-site magnetometers \cite{2013PhRvD..87l3009T}, and based on this measurement, the latter has been estimated in Ref.~\cite{2014PhRvD..90b3013T}, assuming the power-law form of the transfer function for magnetic coupling, $r_{i}(f)$ [see Eq.~(\ref{eq:conv_noise}) or (\ref{eq:snrBB2})]. Though the measurement results are not the direct observations through the interferometric signals, it is still worth comparing their results, at least, to validate or justify the simplification and assumptions made in the analytic model.


For a quantitative comparison, we first need to incorporate the effects arising from nonidealistic situations into the model predictions. Unlike the cases considered in previous section, the observed Schumann resonance peak appears as a narrow-band sharp distribution over the frequencies around the slightly shifted eigenfrequency $f'_\ell\simeq 0.78\,f_\ell$. This is basically due to the imperfect conductivity at the boundary as well as the effective dielectric property inside the Earth-ionosphere cavity system. In order to account for this, we introduce the line shape function given by \cite{1998clel.book.....J}
\begin{align}
|E_\ell(f)|^{2} \propto \frac{1}{(f-f'_{\ell})^{2}+\{f_{\ell}/(2\mathcal{Q})\}^{2}}, 
\label{eq:spectral_form}
\end{align}
where the quantity $\mathcal{Q}$ is the so-called quality factor, which is defined by the ratio of the resonant frequency to the full width at half maximum. We set $\mathcal{Q}=5$, which is close to the one inferred from the observed spectrum of Schumann resonances (e.g., Refs.~\cite{2013PhRvD..87l3009T,1998clel.book.....J,1983JATP...45...55S,2007quality}). 
 Convolving with this line shape function, the correlated noise, described originally as the sum of discrete modes [see Eq.~(\ref{eq:SB_discrete})], may be replaced with the one in Eq.~(\ref{eq:snrBB2}): 
\begin{align}
\langle S _{\rm B} \rangle 
&= T\, \int_{0}^{\infty}\,df \,\,\mbox{Re}\,\bigl[r_{1}^{\ast}(f)\,r_{2}(f)\,\bigr]\,M_{1 2}(f)\,\tilde{Q}(f),
\nonumber
\end{align}
where the magnetic noise spectrum $M_{12}$ is now expressed as (with an appropriate normalization)
\begin{equation}
 M_{1 2}(f) =\frac{1}{8\pi}P_{\rm B}(f)\,\,\sum_{\ell} \frac{|E_\ell(f)|^{2}}{|E_\ell(f'_{\ell})|^{2}}\,
\gamma_{\ell}^{\rm B}(\widehat{\bm r}_{1},\widehat{\bm r}_{2}).
\label{eq:mij}
\end{equation}
In the above, the quantity $P_{\rm B}$ is the power spectral density of the 
magnetic fields [see Eq.~(\ref{eq:expectation_valueB}) for definition in the 
discrete case]. It is known that the amplitude of magnetic field is typically a 
few pT\,\,Hz$^{-1/2}$ and the spectrum is approximately described by a power-law at the frequencies of the Schumann resonances \cite{Ryan:2014,1990PhDT........90C}. 
Based on the discussion in Ref.~\cite{Allen:1997ad}, we adopt the following power-law spectrum: 
\begin{align}
 P_{\rm B}(f)=A \left(\frac{f}{10 {\rm Hz}}\right)^{-0.88},
\label{eq:mg_spectrum}
\end{align}
with the amplitude being set to $A^{1/2}=5.89$\,pT\,\,Hz$^{-1/2}$, to closely match the 
recent measurement \cite{Ryan:2014}
\footnote{The amplitude $A^{1/2}$ set here seems somewhat larger than the measured value of Ref.~\cite{Ryan:2014}. 
But, our definition of power spectrum $P_{\rm B}$ is not directly compared to the measured value. It should be multiplied by the factor $(8\pi)^{-1/2}$. Then, the values $(P_{\rm B}/(8\pi))^{1/2}$ 
at the first two resonance peaks ($f=7.82$ and $13.96$ Hz) become 1.31 and 1.02 pT\,\,Hz$^{-1/2}$,
respectively, which are indeed consistent with the measured results in Ref.~\cite{Ryan:2014}.}. 
Then, for a given detector setup listed in 
Table \ref{position_separation}, the functional form of the magnetic spectrum 
$M_{12}$ is uniquely determined with a help of the analytic expression for 
$\gamma_\ell^{\rm B}$ [Eq.~(\ref{eq:Gamma}) or (\ref{eq:gamma_analytic})] if we 
further specify the unit vector $\widehat{\bm X}_i$ at each detector. In what 
follows, assuming that $\widehat{\bm X}_i$ is independent of the frequency, we characterize it by the orientation angle, $\psi_i$, defined counterclockwise from the local east direction on the tangent plane at each detector.

\begin{figure}[tb]
\hspace*{-0.8cm}
\includegraphics[width=9cm,angle=0,clip]{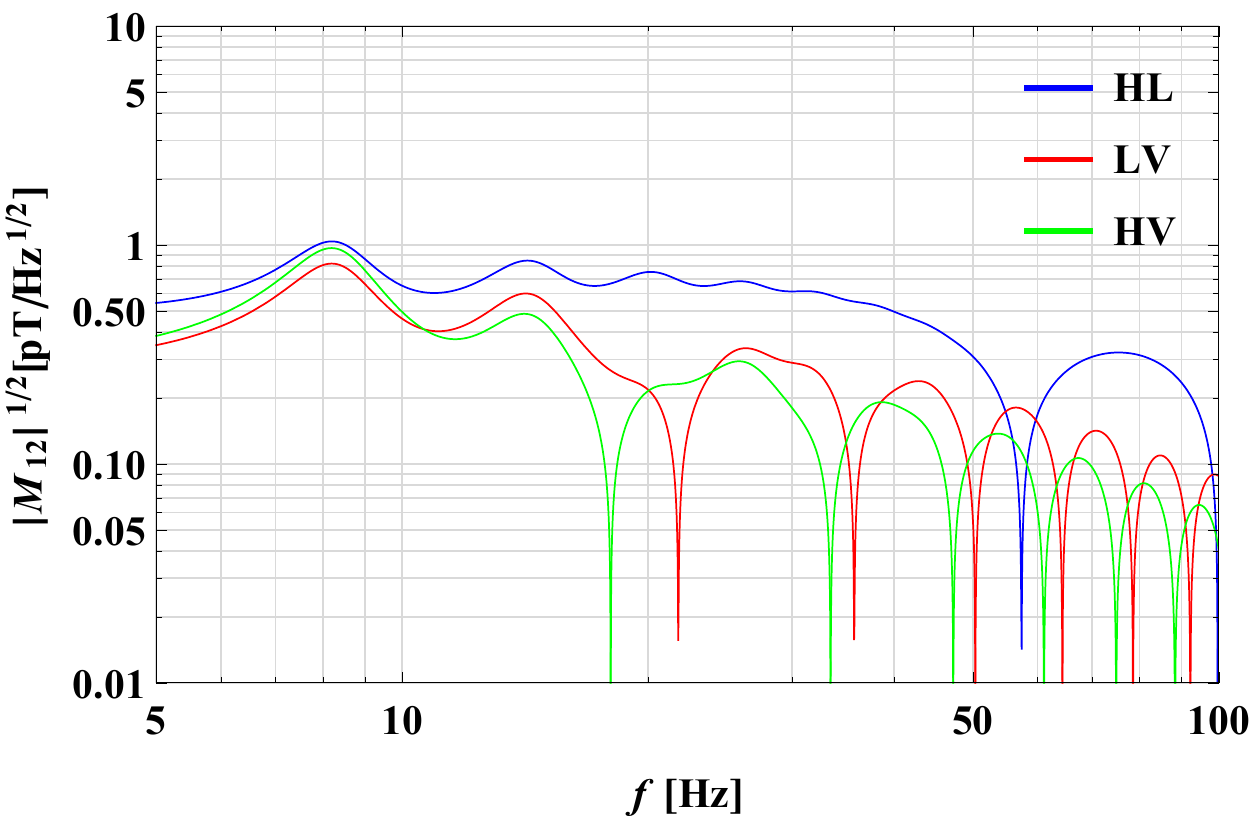}
\caption{Square root of the magnetic noise power spectrum, $|M_{12}|^{1/2}$ as function of frequency for the pairs of HL (blue), LV (red) and HV (green). In plotting these results, the parameters characterizing the orientation of projection vector in analytic model, $\psi_i$, are chosen so as to reproduce the behaviors seen in Fig.~2 of Ref.~\cite{2013PhRvD..87l3009T}.
\label{fig:MHV_MLV}
}
\end{figure}

\subsection{Magnetic noise power spectrum $M_{12}$}
\label{subsec:comparison_M_12}

Fig.~\ref{fig:MHV_MLV} presents the analytic prediction of the magnetic noise spectrum $M_{12}$ for the pairs among LIGO Hanford (H), Livingston (L), and Virgo (V) detectors. To be precise, what is shown here is the square root of the magnetic noise spectrum, $|M_{12}|^{1/2}$, for the HL (blue), LV (red), and HV (green) pairs. In plotting the results, the orientation angle $\psi_i$ characterizing the projection vector has been chosen such that the results reasonably reproduce those measured during LIGO S6 runs\footnote{For the HL pair, Ref.~\cite{2013PhRvD..87l3009T} presents two other measurements called S5 HL A and S5 HL B, which slightly differ from S6 HL. A plausible reason for this comes from the seasonal variation of the Schumann resonances, or different setup of the magnetometers.}, as shown in Fig.~2 of Ref.~\cite{2013PhRvD..87l3009T}.

In Fig.~\ref{fig:MHV_MLV}, different oscillatory features for the three lines basically come from the differences of the geometric distance between each detector pair. This has been already seen in Fig.~\ref{fig:normalized_gamma}. Varying $\psi_i$, the overall amplitude of the spectrum is changed, and the oscillating phase is slightly shifted. With an appropriate choice of $\psi_i$, one can qualitatively recover the measurement results of Ref.~\cite{2013PhRvD..87l3009T}, shown in their Fig.~2. The derived values of the parameter $\psi_i$ are $\psi_i=1.7^\circ$ (H), $105.1^\circ$ (L), and $132.1^\circ$ (V). Although the agreement is still at qualitative level, we think it remarkable in the sense that despite several simplifications, the model successfully describes the major trend of the measurement results, and this can be achieved by adjusting only the three parameters, $\psi_i$.

\begin{table}[b]
\begin{ruledtabular}
\caption{Absolute values of signal-to-noise ratio, $|\rm SNR_{B}|$, for HL pair, assuming the one-year observation. The second column describes the coupling parameters of the length degree of freedom and the angular degree of freedom assuming beam offsets of 1 and 3 mm. The fourth column represents the derived results based on the analytic model, which are compared with those obtained by Ref.~\cite{2014PhRvD..90b3013T} (rightmost column, see also their Table II).}
\newcolumntype{C}{>{\centering}p{5em}}
\begin{tabular}{CCC|C|C}
Coupling  & ($\kappa$, $b$) & Spectral \\ index $\alpha$ & $|\rm SNR_{B}|$ \\ This work & $|\rm SNR_{B}|$ \\ Ref.~\cite{2014PhRvD..90b3013T} \tabularnewline
\hline
\hline
Angular \\ (1mm) & (0.25, 1.74) & 2/3 \\ 0 \\ -2 & 26 \\ 27 \\ 24 & 29 \\ 30 \\ 24 \tabularnewline
Angular \\ (3mm) & (0.75, 1.74) & 2/3 \\ 0 \\ -2 & 231 \\ 244 \\ 219 & 260 \\ 270 \\ 220  
\tabularnewline 
Length           & (2, 2.67) & 2/3 \\ 0 \\ -2 & 317 \\ 382 \\ 497 & 330 \\ 380 \\ 470 
\tabularnewline
\end{tabular}
\label{tab:snrB_HL}
\end{ruledtabular}
\end{table}

\subsection{Signal-to-noise ratio ${\rm SNR}_{\rm B}$}
\label{subsec:comparison_SNR_B}

\begin{figure}[tb]
\hspace*{-0.5cm}
\includegraphics[width=9.cm,angle=0,clip]{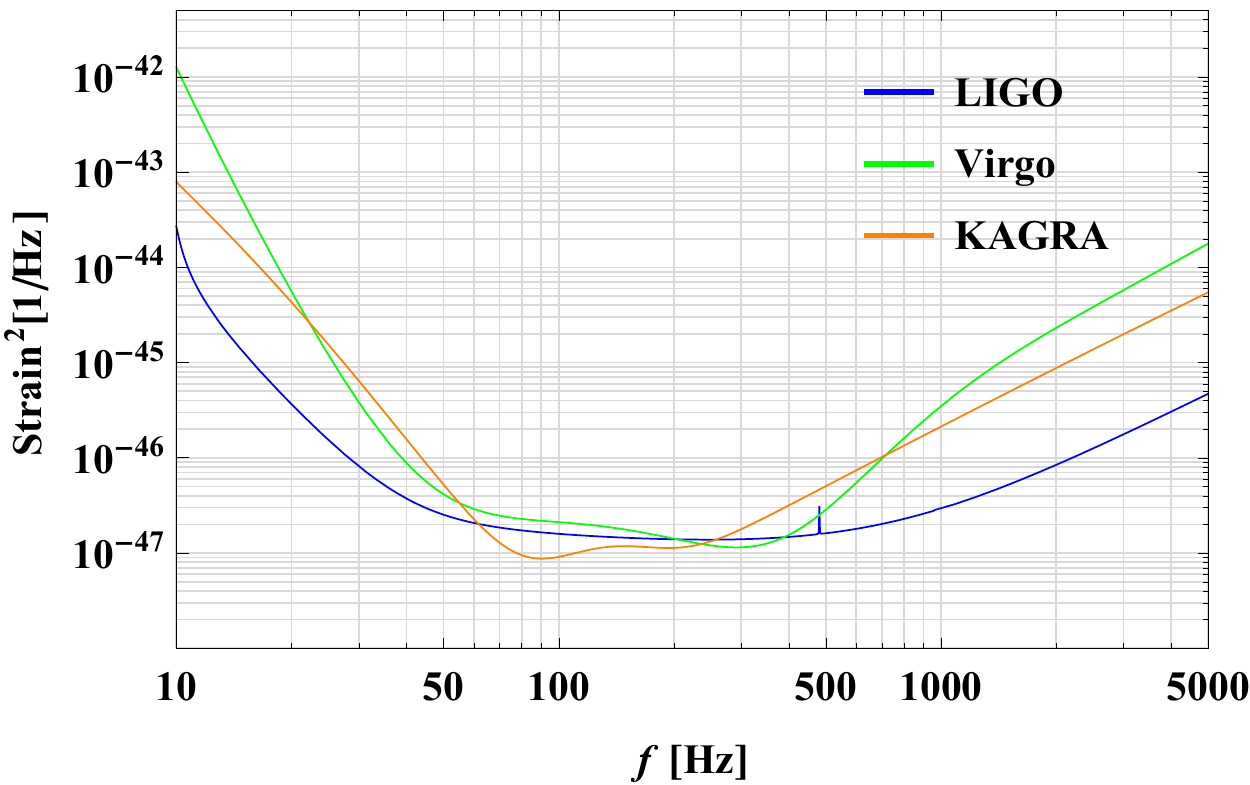}\\
\caption{Instrumental noise spectrum, $P_i(f)$, for second-generation detectors used for the analysis in Sec.~\ref{subsec:comparison_SNR_B} and \ref{sec:implication}: LIGO (blue), Virgo (green), KAGRA (orange). Note that in the analysis, we assume that the noise spectral density in LIGO Hanford and Livingston, as well as LIGO India, are identical.}
\label{fig:noise_curve}
\end{figure}

Having confirmed that the model describes the major trend of the measured magnetic noise spectra, let us next consider the signal-to-noise ratio, ${\rm SNR}_{\rm B}\equiv\langle S_{\rm B}\rangle/\sigma$, and compare the model predictions with those obtained by Ref.~\cite{2014PhRvD..90b3013T}. To be precise, the quantity to be estimated is the ratio of the expectation value of the cross-correlation statistic, $S$, undesirably dominated by the correlated magnetic noise, to the dispersion of $S$, which is determined by the auto-correlation of the instrumental noises, adopting the optimal filter given at Eq.~(\ref{eq:filter_function}). Thus, on top of the magnetic noise power spectrum computed above, we need to know the instrumental noise spectrum $P_i(f)$ and overlap reduction function $\gamma_{12}^{\rm G}$ for each detector (pair). Here, for the noise spectrum, we use the table of numerical data for LIGO Hanford/Livingston detectors \cite{Shoemaker:2014} (see Fig.~\ref{fig:noise_curve}). 
The expression of the overlap reduction function is analytically known (e.g., Refs.~\cite{Allen:1997ad,Flanagan:1993ix}), and we use the one summarized in Appendix \ref{ap:overlap}, together with the geometric information of the GW detectors listed in Table \ref{position_separation}.  

Then, provided the transfer function of the magnetic noise coupling, $r_{i}(f)$, the impact of correlated magnetic noise is evaluated under the assumption of the  expected stochastic GW signal, $\Omega_{\rm gw}$, for which we assume the power-law shape, $\Omega_{\rm gw}\propto f^\alpha$. Table \ref{tab:snrB_HL} summarizes the estimated results of $|{\rm SNR}_{\rm B}|$ for one-year observation $({\rm i.e.,}\, T=1\,{\rm yr})$, based on the following power-law form of the transfer function:
\begin{align}
r_{i}(f)=\kappa  \times 10^{-23}\left(\frac{f}{10 \,{\rm Hz}}\right)^{-b} [{\rm strain}\,\,{\rm pT}^{-1}]\,.
\label{eq:coupling}
\end{align}
Here, we use the same orientation angle $\psi_i$ as derived in Sec.~\ref{subsec:comparison_M_12} (see Fig.~\ref{fig:MHV_MLV}).  Following Ref.~\cite{2014PhRvD..90b3013T}, in Table \ref{tab:snrB_HL}, the absolute values of ${\rm SNR}_{\rm B}$ are presented for various set of parameters $(\kappa, b)$, which are determined experimentally according to the mechanism of magnetic coupling. Possible scenarios for magnetic couplings we consider are the length coupling, which is a direct coupling to the length degree of freedom for mirror motion, and the angular coupling as an indirect coupling through the angular motion. We set the parameters $(\kappa, b)$ to $(2, 2.67)$ for the length coupling, and for the angular coupling, we adopt  $(0.25, 1.74)$  and $(0.75, 1.74)$ for 1 and 3 mm beam offsets, respectively (See Ref.~\cite{2014PhRvD..90b3013T} for detail).

\begin{table}[tb]
\begin{ruledtabular}
\caption{Same as in Table \ref{tab:snrB_HL}, but for HV and LV pairs. }
\newcolumntype{C}{>{\centering}p{5em}}
\begin{tabular}{CCC|CC}
Coupling  & ($\kappa$, $b$) & Spectral \\ index $\alpha$ & $|\rm SNR_{B}|$ \\ for HV & $|\rm SNR_{B}|$ \\ for LV \tabularnewline
\hline
Angular \\ (1mm) & (0.25, 1.74) & 2/3 \\ 0 \\ -2 & 0.75 \\ 0.86 \\ 0.64 & 1.2 \\ 1.4 \\ 1.0 \tabularnewline
Angular \\ (3mm) & (0.75, 1.74) & 2/3 \\ 0 \\ -2 & 6.8 \\ 7.8 \\ 5.7 & 11 \\ 12 \\ 9.3  
\tabularnewline 
Length           & (2, 2.67) & 2/3 \\ 0 \\ -2 & 5.2 \\ 6.0 \\ 1.8 & 7.3 \\ 8.2 \\ 4.5 
\tabularnewline
\end{tabular}
\label{tab:snrB_HV_LV}
\end{ruledtabular}
\end{table}

Quite remarkably, the derived signal-to-noise ratios (fourth column) quantitatively match those obtained by Ref.~\cite{2014PhRvD..90b3013T}. Since the estimated values of Ref.~\cite{2014PhRvD..90b3013T} are based on the measured results of the magnetic noise spectrum with which the analytic model only provides a qualitative agreement, this is a considerable success. Rather, the results imply that a detailed magnetic field structure or a more complicated and realistic setup is not essential to estimate the impact of correlated noise, and even the simplified analytic model can give a quantitative estimate.

As an illustration, we repeat the same analysis as given above for other detector pairs (i.e., HV and LV pairs), and summarize the results in Table \ref{tab:snrB_HV_LV}. Here, we used the fitting form of the noise spectrum for Virgo, presented in Ref.~\cite{Manzotti:2012uw}. Again, we assumed the observation time, $T=1\,{\rm yr}$. Then the predicted values of signal-to-noise ratio for HV and LV pairs are rather smaller than those obtained for HL pair. The main reason for this is basically ascribed to the detector sensitivity of Virgo and magnetic noise spectra seen in Fig.~\ref{fig:MHV_MLV}. That is, at the low-frequency band around $f\lesssim50$\,Hz, where the detector is most sensitive to the stochastic GW signal, the LIGO detectors have a better sensitivity than Virgo (see Fig.~\ref{fig:noise_curve}), and the magnetic noise spectrum $M_{12}$ of HL pair has a larger amplitude with slowly oscillatory behavior. These two facts result in a large impact of the correlated magnetic noise for the HL pair, while the impacts are reduced to some extent for HV and LV pair. Nevertheless, even for HV and LV pairs, the amplitude of correlated magnetic noise is still large, and can have a potential to exceed the stochastic GW signal. In Table \ref{tab:omega_gw_HLV}, assuming the flat spectrum of stochastic GWs and adopting the length coupling of transfer function, we convert the impact of the correlated noise into the amplitude of stochastic GW, and derive $\Omega_{\rm gw}h^{2}$ by setting ${\rm SNR}_{\rm G}=|{\rm SNR}_{\rm B}|$. The resultant amplitude $\Omega_{\rm gw}h^2$, presented in upper-right part, is indeed comparable or slightly larger than the minimum detectable amplitude for GW signals (lower-left part), which is derived assuming ${\rm SNR}_{\rm G}=5$ and $T=1\,{\rm yr}$.

\begin{table}[tb]
\begin{ruledtabular}
\caption{Upper right: $\Omega_{\rm gw}h^{2}$ satisfying $\rm{SNR_{G}=|SNR_{B}}|$ for HL, HV and LV pairs
in which we assume that stochastic GWs have the flat spectrum and adopt the length couping as a transfer function.
Lower left: $\Omega_{\rm gw}h^{2}$ satisfying $\rm{SNR_{G}=5}$ for the observation time $T=1 {\rm yr}$.}
 \begin{tabular}{l ccc} 
  & H   & L & V \\ \hline \hline
  H \,\,\, & $\ast$ & $1.2 \times 10^{-7}$ & $2.0 \times 10^{-8}$ \\
    L & $1.6 \times 10^{-9}$ & $\ast$ & $2.3 \times 10^{-8}$ \\ 
   V & $1.6 \times 10^{-8}$ & $1.4 \times 10^{-8}$ & $\ast$ \\ 
  \end{tabular}
\label{tab:omega_gw_HLV}
\end{ruledtabular}
\end{table}

\begin{figure*}[tb]
\hspace*{-0.5cm}
\includegraphics[width=6cm,angle=0,clip]{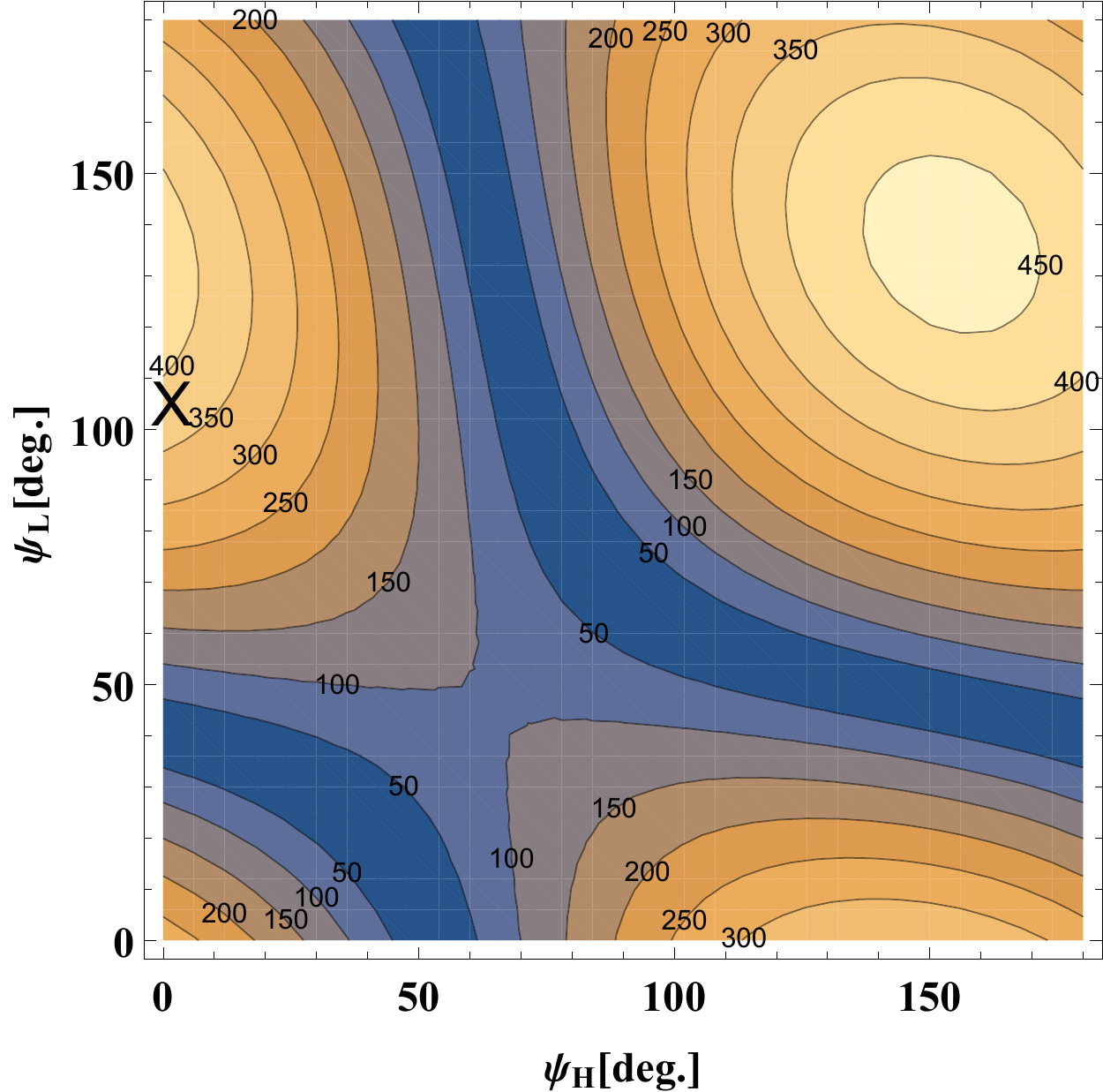}
\includegraphics[width=6cm,angle=0,clip]{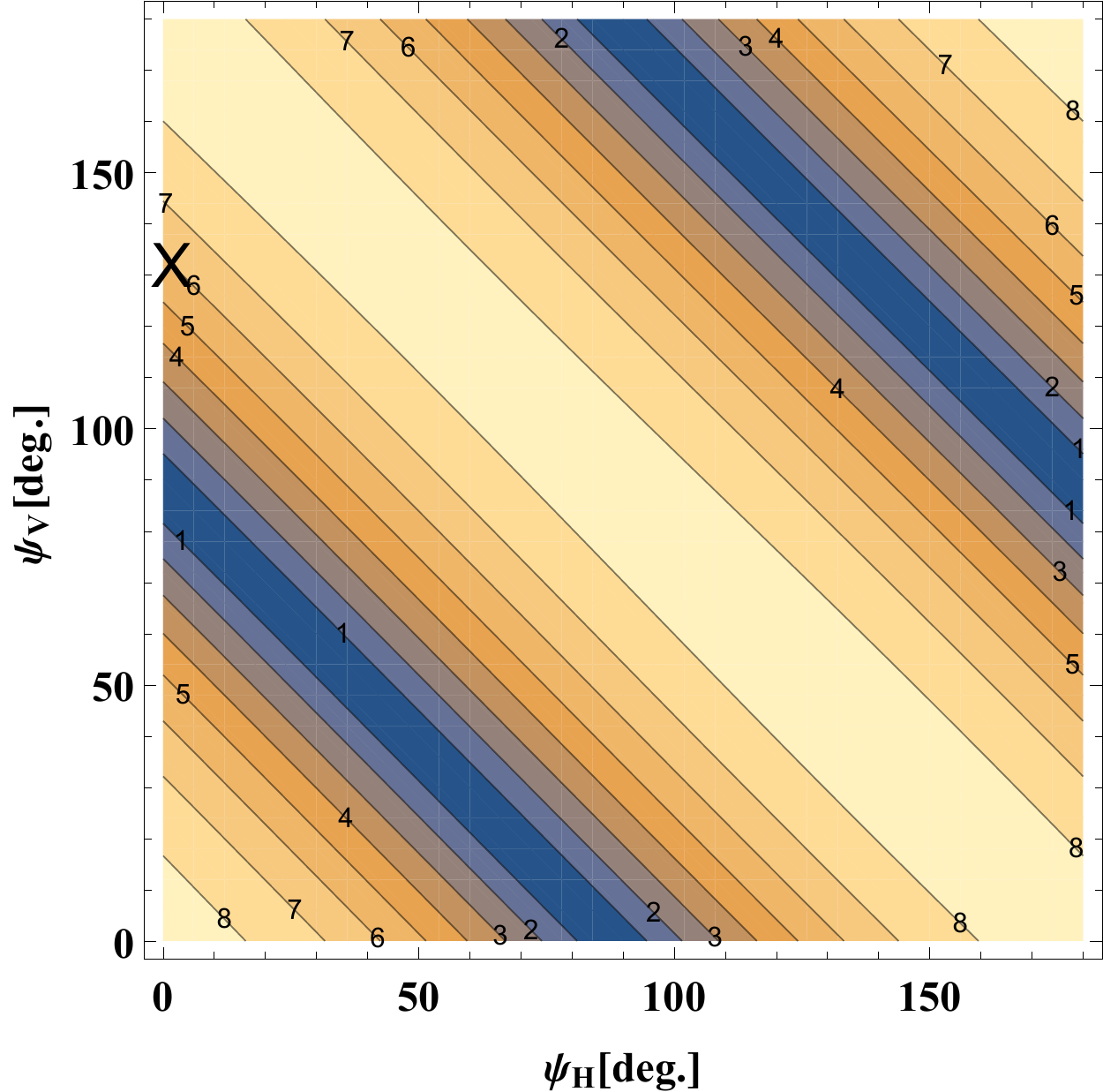}
\includegraphics[width=6cm,angle=0,clip]{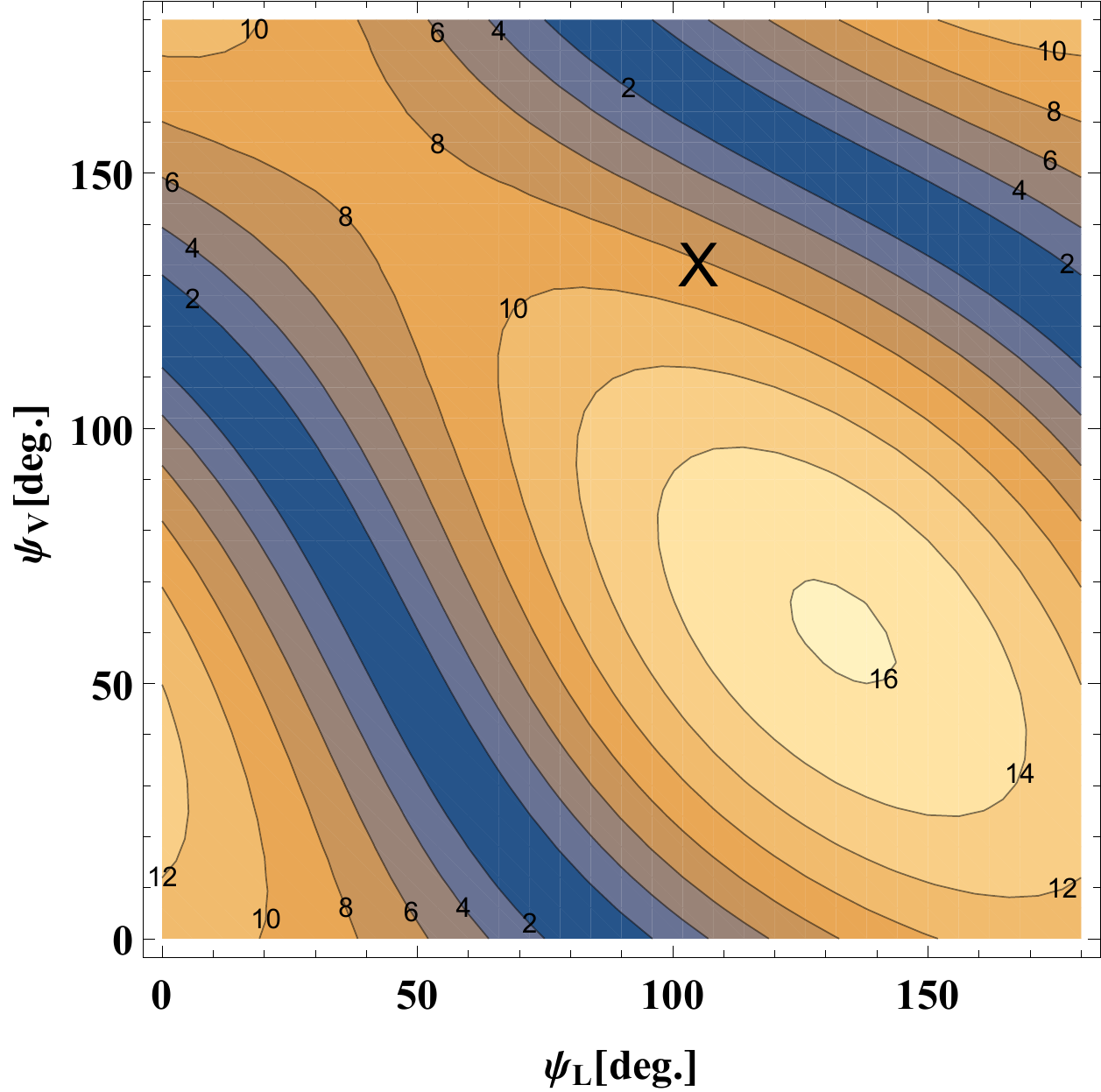}
\caption{Two dimensional contours of $|{\rm SNR_{B}}|$ as function of $\psi_{1}$ and $\psi_{2}$ for HL(left), HV(middle) and LV(right) pairs, assuming the one year observation and flat spectral shape for the expected stochastic GWs. In computing $|{\rm SNR_{B}}|$, we adopt the transfer function in Eq.~(\ref{eq:coupling}), with the parameters of length coupling, $(\kappa, b)=(2, 2,67)$ (see Table \ref{tab:snrB_HL}).
Each black cross symbol is marked at $(\psi_{\rm H}, \psi_{\rm L}, \psi_{\rm V})=(1.7, 105.1, 132.1)$ in units of degree. The values of $|{\rm SNR_{B}}|$ at black cross symbol correspond to the one for the length coupling in Tables \ref{tab:snrB_HL} and \ref{tab:snrB_HV_LV}.}
\label{fig:SNR_B_as_func_psi}
\end{figure*}

\section{Implications}
\label{sec:implication}

In this section, based on the analytic model, we extend the analysis in Sec.~\ref{sec:comparison}, and explore the potential impact of correlated magnetic noises including the upcoming second-generation detectors, KAGRA and LIGO India.

To investigate its impact in more generic way, we first recall that the impact 
of correlated noise depends crucially not only on the geometrical configuration 
of the detector pair, but also on how one can shield the magnetic fields and 
mitigate the coupling of mirror control system with magnetic fields. In the 
present analytic model, the latter is described by the transfer function 
$r_i(f)$ and projection vector $\widehat{\bm X}_i$ or the orientation angle $\psi_i$ introduced at Sec.~\ref{sec:comparison}. While we found a reasonable set of parameters $\psi_i$ for LIGO Hanford/Livingston and Virgo that reproduce the results in Refs.~\cite{2013PhRvD..87l3009T,2014PhRvD..90b3013T}, we note cautiously that their results are based on the measurements by magnetometer that monitors the local environment around GW detectors. Thus, the parameters might differ from the one derived from the interferometric signals. In addition, the upcoming detectors, KAGRA and LIGO India, are uncertain for the detail of the magnetic coupling until the construction is completed\footnote{At KAGRA site, the magnetic fields are kept monitored during the construction phase, and a global correlation of the magnetic field with other detector site have been detected.}.

Here, we shall below assume that all the detectors have the same transfer function as given in Eq.~(\ref{eq:coupling}), adopting the parameters of the length coupling given by $(\kappa,\,b)=(2,\,2.67)$. Then, the remaining uncertainty of the magnetic coupling is the projection vector characterized by the orientation angle $\psi_i$. Let us see how the parameter $\psi$ changes the impact of correlated noises.

Fig.~\ref{fig:SNR_B_as_func_psi} shows the dependence of ${\rm SNR}_{\rm B}$ defined in Sec.~\ref{sec:comparison} [Eq.~(\ref{eq:def_SNR_B})] on the parameter $\psi_i$ for HL (left), HV (middle), and LV (right) pairs, assuming the one-year observations and flat spectral shape for the expected stochastic GWs. To be precise, what is plotting is the absolute value, $|{\rm SNR}_{\rm B}|$, and since it is given as the function of angle $\psi_i$ modulo $\pi$ [see e.g., Eq.~(\ref{eq:gamma_analytic})], the results are shown in the range, $0^\circ\leq\psi_i\leq180^\circ$. In each panel, the specific values of $\psi_i$ used to derive the results in Fig.~\ref{fig:MHV_MLV}, Tables \ref{tab:snrB_HL} and \ref{tab:snrB_HV_LV} are indicated as the black cross symbol. From Fig.~\ref{fig:SNR_B_as_func_psi}, we see that the parameters examined in previous section give a moderately large value of $|{\rm SNR}_{\rm B}|$, but the worst case has even more large. For the possible worst cases, setting $|{\rm SNR}_{\rm B}|={\rm SNR}_{\rm G}$, the impact of correlated magnetic noises is estimated to give $\Omega_{\rm gw}h^2=1.5\times10^{-7}$ (HL), $2.8\times10^{-8}$ (HV), and $4.6\times10^{-8}$ (LV).  

\begin{figure}[tb]
\includegraphics[width=8.5cm,angle=0,clip]{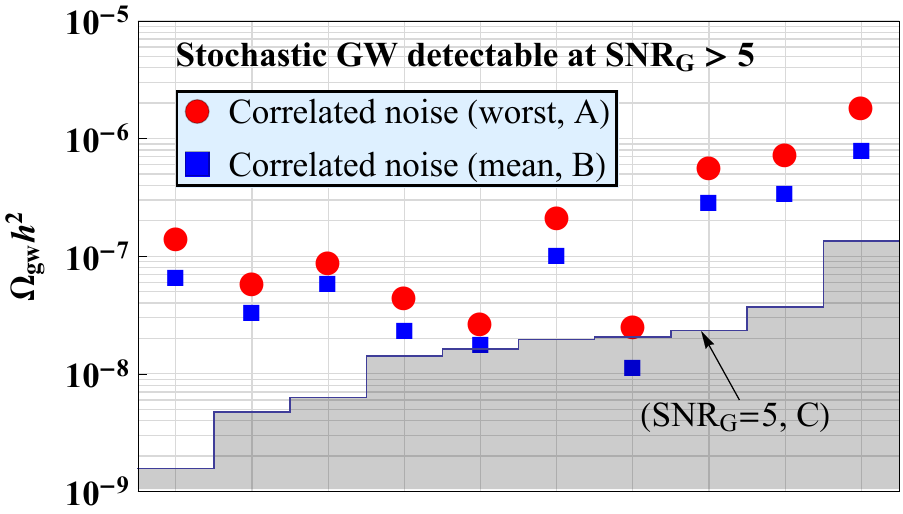}\\
\vspace{-2mm}
\hspace{0.15cm}
\includegraphics[width=8.4cm,angle=0,clip]{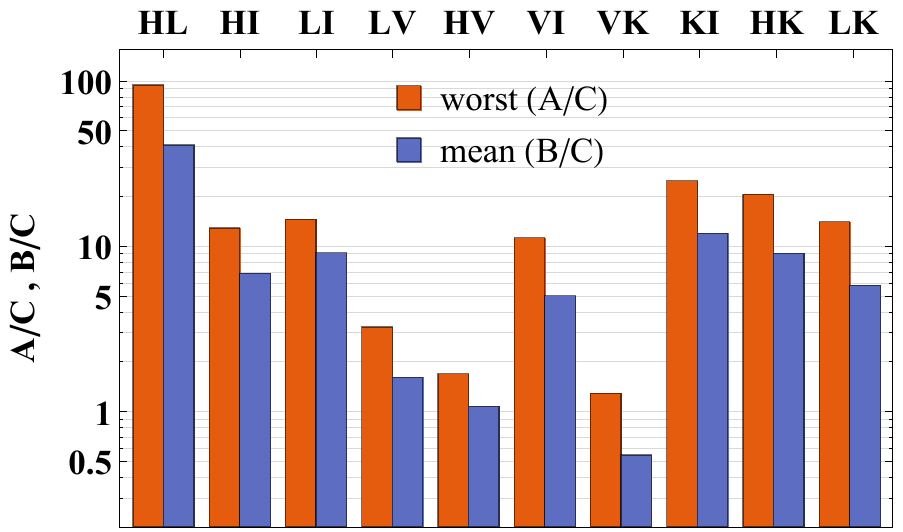}\\
\caption{Potential impact of the correlated magnetic noise on the detection of stochastic GWs. For each detector pair, varying the orientation angles $\psi_i$, we first look for the mean and maximum value of $\rm{|SNR_B|}$. The resultant values are then translated to $\Omega_{\rm gw} h^2$ by setting $\rm{|SNR_B|=SNR_G}$. Upper panel shows the estimated results of $\Omega_{\rm gw} h^2$ for worst (filled circles) and mean (filled squares) cases (respectively labeled as A and B). Note that we adopt the transfer function in Eq.~(\ref{eq:coupling}), with the parameters of length coupling. 
For reference, assuming the one-year observation and flat spectrum, amplitudes of stochastic GWs detectable at $\rm{SNR_G}>5$ and $<5$ are shown in nonshaded and shaded regions, respectively.  The boundary between the two, depicted as a stepwise line, corresponds to the minimum detectable GW amplitude of $\rm{SNR_G}=5$ (labeled as C).   In lower panel, to estimate the relative impact of the correlated magnetic noise, the amplitudes of  $\Omega_{\rm gw} h^2$ for worst and mean cases are normalized by the minimum detectable GW amplitude of $\rm{SNR_G}=5$ for each pair, and the results are respectively plotted as red and blue histograms (i.e,. $\rm{A/C}$ and $\rm{B/C}$).}
\label{fig:comparison_Omega_gw}
\end{figure}

\begin{table}[tb]
\begin{ruledtabular}
\caption{Equivalent to the worst and mean values in the upper panel in Fig.~\ref{fig:comparison_Omega_gw}.
Upper right: $\Omega_{\rm gw} h^{2}$  satisfying that $\rm{SNR_{G}}$ is equal to $|\rm{SNR_{B}}|$ of the worst case. 
Lower left: $\Omega_{\rm gw} h^{2}$ satisfying that $\rm{SNR_{G}}$ is equal to the mean value of $\rm{|SNR_{B}|}$. Observation time and transfer function are same in Table \ref{tab:omega_gw_HLV}.}
 \begin{tabular}{l ccccc} 
& H                    & L                    & V                    & K                    & I \\ \hline \hline
  H \,\,\, & $\ast$               & $1.5 \times 10^{-7}$ & $2.8 \times 10^{-8}$ & $7.6 \times 10^{-7}$ & $6.1 \times 10^{-8}$\\
         L & $6.4 \times 10^{-8}$ & $\ast$               & $4.6 \times 10^{-8}$ & $1.9 \times 10^{-6}$ & $9.1 \times 10^{-8}$\\ 
         V & $1.7 \times 10^{-8}$ & $2.3 \times 10^{-8}$ & $\ast$               & $2.6 \times 10^{-8}$ & $2.2 \times 10^{-7}$\\
         K & $3.3 \times 10^{-7}$ & $7.8 \times 10^{-7}$ & $1.1 \times 10^{-8}$ & $\ast$               & $5.8 \times 10^{-7}$\\
         I & $3.3 \times 10^{-8}$ & $5.7 \times 10^{-8}$ & $9.8 \times 10^{-8}$ & $2.8 \times 10^{-7}$ & $\ast$ 
  \end{tabular}
\label{tab:omega_gw_all}
\end{ruledtabular}
\end{table}

In Fig.~\ref{fig:comparison_Omega_gw}, similar analysis is performed for all the pairs of detectors including KAGRA and LIGO India, and the potential impact of the correlated magnetic noise is quantified as $\Omega_{\rm gw}h^2$. In estimating the impact for LIGO India, we use the same instrumental noise curve as used in LIGO Hanford/Livingston. As for the noise spectrum of KAGRA, we use the fitting form summarized in Ref.~\cite{Manzotti:2012uw} (Fig.~\ref{fig:noise_curve}). In the upper panel of Fig.~\ref{fig:comparison_Omega_gw}, filled circles indicate the worst case amongst possible combination of $\psi_i$ for each pair, while filled squares represent the mean value of $\Omega_{\rm gw}h^2$. These are compared with the amplitude of stochastic GWs detectable at ${\rm SNR}_{\rm G}>5$ and $<5$, respectively shown as nonshaded and shaded regions, assuming the one-year observation and flat spectrum. The boundary between the two, depicted as a stepwise line,  corresponds to the minimum detectable amplitude of ${\rm SNR_G}=5$. Here, the results are sorted (intentionally) as increasing the minimum delectable amplitude. The estimated values are also summarized in Table \ref{tab:omega_gw_all}.

Overall, the amplitude of the correlated noise in the worst cases can exceed the amplitude of the detectable GWs by more than one order of magnitude, and it becomes larger as increasing the detectable amplitude of stochastic GWs. Interestingly, however, there are exceptional detector pairs for which the correlated noise amplitude is almost comparable to the detectable amplitude of GWs. These are LIGO Hanford-Virgo (HV) and Virgo-KAGRA (VK) pairs. 
The lower panel of Fig. 7 plots  the relative impact of the correlated magnetic noise on the detection of stochastic GWs, just dividing the worst and mean values of $\Omega_{\rm gw}h^2$  (respectively labeled as A and B) by the GW amplitude detectable at $\mbox{SNR}_{\rm G}=5$ (labeled as C), i.e., A/C and B/C. Clearly, the normalized amplitude is close to unity for HV and VK pairs. The main reason for this is that the integral of $\mbox{SNR}_{\rm B}$ involves the product of oscillating functions, $\gamma_\ell^{\rm B}$ and $\gamma_{12}^{\rm G}$. Due to the detector sensitivity, the integral becomes almost converged when integrating over frequencies up to $f\sim50$\,Hz (see Fig.~\ref{fig:noise_curve}).  Thus, the behavior of the integrand below $50$\,Hz gives a large impact on $\mbox{SNR}_{\rm B}$. Depending on the detector's orientation (characterized by $\delta_{\rm G}$ and $\Delta_{\rm G}$, see Eq.~(\ref{eq:gamma_analytic_G}) and Table~\ref{position_separation}), the integrand exhibits a severe cancellation, and among all pairs, the HV and VK pairs turn out  to have a rather large cancellation at $f\lesssim50$\,Hz.

The results presented in Fig.~\ref{fig:comparison_Omega_gw} assume specific coupling parameters for the transfer function, but recalling the fact that the amplitude  $|\mbox{SNR}_{\rm B}|$ is proportional to  $\kappa^2$ or precisely to $r_i(f)r_j(f)$, it can be translated to the cases with different amplitude of transfer function. Fig.~\ref{fig:comparison_Omega_gw} implies that the coupling with magnetic fields must be mitigated, at least, by a factor of $3\sim10$ (corresponds to the square root of the ratio shown in lower panel of Fig.~\ref{fig:comparison_Omega_gw}) at each detector in order to demonstrate the best performance of the stochastic GW searches. Ref.~\cite{Effler:2015p} recently reported an updated calibration of the coupling function for LIGO, and found that the amplitude of the coupling is substantially reduced by more than 1 order of magnitude in the latest instrumental setup. Still, however, the impact of the correlated magnetic noises can be potentially large for other detectors. In this respect,  the development of the methodology to subtract the correlated noises like those performed in Ref.~\cite{2014PhRvD..90b3013T} would be crucial for future detection of stochastic GWs.

\section{Conclusion}
\label{sec:conclusion}
%
%
The first direct detection of gravitational waves (GW) by the laser interferometer, LIGO, has opened a new window to probe the Universe, and together with the upcoming second-generation detectors, a significant number of GW events will be detected. Increasing the detector sensitivity, we will have a potential to measure the stochastic background of GW, which may give many hints not only on the population of binary sources over the Universe but also on the physics in the early Universe. The detection of stochastic GWs, however, might not be straightforward in the presence of correlated magnetic noises. 

In this paper, we have investigated the impact of correlated magnetic noises on the detection of stochastic GWs via upcoming/ongoing ground-based laser interferometers. In particular, we presented a simple analytic model, which enables us to study how the global magnetic fields formed inside the Earth-ionosphere cavity leads to a nonvanishing correlation between detector pair. Assuming that the global magnetic fields are described by the random superposition of the axisymmetric transverse magnetic (TM) modes formed at the isotropically distributed exciting sources, we have derived the analytic expressions for correlated magnetic noises. The resultant expression includes the quantity, $\gamma_\ell^{\rm B}$, which is analogous to the overlap reduction function $\gamma_{12}^{\rm G}$ that appears in the cross correlation of the GW signals. The crucial point is that the function $\gamma_\ell^{\rm B}$ depends on the geometric configuration of the detector pair through the directional-dependent coupling with magnetic fields. We investigated the basic properties of the function $\gamma_\ell^{\rm B}$, and found that it induces the oscillatory features in the correlated noise spectrum. The function $\gamma_\ell^{\rm B}$ does not vanish even for a widely separated detector pair, and exhibits more rapid oscillation. 

We then quantitatively compared the analytic model to the measured results by Refs.~\cite{2014PhRvD..90b3013T,2013PhRvD..87l3009T}. Incorporating the effect of line-shape function into the analytic model, we have shown that the model reproduces the major trends of the magnetic noise spectrum measured by Ref.~\cite{2013PhRvD..87l3009T} via the magnetometers between LIGO Hanford/Livingston and Virgo. Based on this, the impact of correlated noises on the cross-correlation signal has been also estimated. We found that the fake signal-to-noise ratios induced by the correlated noise match quite well with those obtained by Ref.~\cite{2014PhRvD..90b3013T}. Finally, taking account of the uncertainties in the magnetic coupling with interferometers, we have estimated the impact of correlated noises on the upcoming second-generation detectors, including KAGRA and LIGO India, finding that even in the pessimistic case that most of the detector pairs are completely dominated by the correlated noise, LIGO Hanford-Virgo and Virgo-KAGRA pairs would be possibly less sensitive to the correlated noise, and can achieve the best sensitivity to the stochastic GWs of the amplitude $\Omega_{\rm gw}h^2\sim(2-3)\times10^{-8}$, assuming the transfer function of the magnetic coupling given by Eq.~(\ref{eq:coupling}) with $(\kappa,b)=(2,2.67)$ and the one-year observation.

The analytic model presented here shows several interesting properties, and despite its simplification, the model can be even used to quantitatively estimate the impact of correlated noises. Of course, toward realistic situations, the model still needs to be improved in several aspects. Apart from the nonstationarity of the Schumann resonances, one is to take account of the anisotropies of the spatial distribution of the exciting sources for global magnetic fields.  Another important point would be to refine the magnetic field of Schumann resonances. With these improvements, together with a refined calibration of magnetic coupling, the model can be used for a more quantitative study, and may give a hint or clue to find a way to mitigate or subtract the correlated noises.  

%
%
\begin{acknowledgments}
The authors would like to thank Atsushi Nishizawa and Kazuhiro Hayama for discussions and helpful comments. Numerical computation was partly carried out at the Yukawa Institute Computer Facility. This work was supported in part by MEXT/JSPS KAKENHI Grant Number JP15H05899 and JP16H03977. 
\end{acknowledgments}

\appendix

\section{Overlap reduction function of stochastic GWs, $\gamma_{12}^{\rm G}$}
\label{ap:overlap}

In this Appendix, we summarize the analytic expressions for the overlap reduction function of stochastic GWs, $\gamma_{12}^{\rm G}$, used in the analysis in Secs.~\ref{sec:comparison} and \ref{sec:implication}.

The overlap reduction function is known to sensitively depend on the geometrical configuration of the pair of GW detectors, and for the ground-based laser interferometers having the two equal-length arms with an opening angle of $90^\circ$, the function $\gamma_{12}^{\rm G}$ is characterized by the three angles, $(\beta$, $\sigma_1^{\rm G}, \sigma_2^{\rm G})$ (e.g., Refs.~\cite{Flanagan:1993ix, Seto:2008sr}). Here, the angle $\beta$ is the same quantity as shown in Fig.~\ref{fig:great_circle}, and represents the separation angle between the two detectors. On the other hand, the angle $\sigma_i^{\rm G}$ characterizes the orientation of each detector, as defined similarly to $\sigma_i$ in Sec.~\ref{property}. To be precise, $\sigma_1^{\rm G}$ ($\sigma_2^{\rm G}$) represents the relative orientation of the bisector of two laser arms of the detector $1$ ($2$, respectively) measured in counterclockwise manner relative to the great circle connecting the two detectors. For convenience, we define $\delta_{\rm G} \equiv (\sigma_{1}^{\rm G}-\sigma_{2}^{\rm G})/2$ and $\Delta_{\rm G} \equiv (\sigma_{1}^{\rm G}+\sigma_{2}^{\rm G})/2$. Then, the overlap reduction function $\gamma_{12}^{\rm G}$, given as function of frequency $f$, is expressed as follows (e.g., \cite{Flanagan:1993ix, Seto:2008sr}): 
\begin{align}
\gamma_{1 2}^{\rm G}(f)=\Theta_{1}(y,\beta)\,\cos(4\delta_{\rm G}) 
+\Theta_{2}(y,\beta)\,\cos(4\Delta_{\rm G}).
\label{eq:gamma_analytic_G}
\end{align}
Here, the functions $\Theta_{1}$ and $\Theta_{2}$ are, respectively, given by 
\begin{align}
& \Theta_{1}(y,\beta)
=\left(j_{0}+\frac{5}{7}j_{2}+\frac{3}{112}j_{4}\right)\cos^{4}\left(\frac{\beta}{2}\right)\,,
\label{eq:theta1}
\\
& \Theta_{2}(y,\beta)
=\left(-\frac{3}{8}j_{0}+\frac{45}{56}j_{2}-\frac{169}{896}j_{4}\right)
\nonumber\\
&\qquad\qquad+\left(\frac{1}{2}j_{0}-\frac{5}{7}j_{2}-\frac{27}{224}j_{4}\right)\cos \beta
\nonumber\\
&\qquad\qquad+\left(-\frac{1}{8}j_{0}-\frac{5}{56}j_{2}-\frac{3}{896}j_{4}\right)\cos(2\beta).
\label{eq:_theta2}
\end{align}
In the above, the function $j_{n}$ is the $n$-th spherical Bessel function of the first kind with its argument given by
\begin{align}
 y \equiv \frac{4\pi f R_{\oplus}}{c}\sin\left(\frac{\beta}{2}\right)\,.
\end{align}

These analytic expressions are used in Sec.~\ref{sec:comparison} and \ref{sec:implication}, together with the angle parameters for specific GW detectors, which are listed in Table \ref{position_separation}.

\begin{table*}[tb]
\begin{ruledtabular}
\caption{Geometrical information of the five detectors and possible pairs made from them (see e.g., Ref.~\cite{Seto:2008sr} for H, L, V and K, and Ref.~\cite{Schutz:2011tw} for I). Diagonal: positions ($\theta$, $\phi$) in a spherical coordinate system. 
The north pole is set by $\theta=0^{\circ}$ and $\phi$ is longitude. 
Upper right: angle parameters ($\delta_{\rm G}$, $\Delta_{\rm G}$) for each pair of detectors.
Lower left: separation angle $\beta$ for each pair of detectors. All parameters presented below are in units of degree.}
\begin{tabular}{lccccc} 
     &  H  & L & V & K & I \\ \hline 
     LIGO Hanford (H)  &(43.5\,, -119.4)&45.3, 62.2&61.1\footnote{We found a minor typo in Ref.~\cite{Seto:2008sr}, and 
corrected it.}, 55.1&89.1, 25.6&74.0, 84.0\\
    LIGO Livingston (L) &27.2&(59.4\,, -90.8)&26.7, 83.1&42.4, 68.1&49.0, 32.3\\ 
    Virgo (V) &79.6&76.8&(46.4\,, 10.5)&28.9, 5.6&40.0, 80.1\\
KAGRA (K) &72.4&99.2&86.6&(53.6\,, 137.3)&51.2, 19.6\\
   LIGO India (I) &113.3&128.2&58.0&57.6&(70.9\,, 74.0)  
  \end{tabular}
\label{position_separation}
\end{ruledtabular}
\end{table*}

\section{Derivation of Eq.~(\ref{eq:Gamma})}
\label{analytic_gamma}

In this Appendix, we derive the analytical expressions of $\gamma_\ell^{\rm B}$, given at Eq.~(\ref{eq:Gamma}) with Eqs.~(\ref{eq:fl}) and (\ref{eq:gl}).

Let us first rewrite the expression given at Eq.~(\ref{eq:gamma_integral}) with the tonsorial form. Recalling that the unit vector $\widehat{\bm e}_i$ is defined by Eq.~(\ref{eq:unit_vect_ei}), we have
\begin{align}
&\gamma_{\ell}^{\rm B}(\widehat{\bm r}_{1},\widehat{\bm r}_{2}) =
\frac{(2\ell+1)(\ell-1)!}{2\pi(\ell+1)!} \nonumber \\
&\qquad\qquad~\times 
\,\Gamma^{b e}(\ell, \widehat{\bm r}_{1}, \widehat{\bm r}_{2})\, \epsilon_{abc}\,
\epsilon_{def}\,{\widehat r}_{1}^{c}\,{\widehat r}_{2}^{f}\,{\widehat X}_{1}^{a}\,{\widehat X}_{2}^{d}\,,
\label{gamma1}
\end{align}
where $\epsilon_{abc}$ is three-dimensional Levi-Civita symbol (permutation tensor), 
and we follow Einstein's summation convention. Here, we defined the matrix $\Gamma^{ab}$: 
\begin{align}
\Gamma^{a b}(\ell, \widehat{\bm r}_{1}, \widehat{\bm r}_{2}) 
= \int_{S^{2}} \,d^2 \widehat{\bm \Omega}\, \frac{\mathcal{P}_{\ell}^{1}(\widehat{\bm \Omega} \cdot \widehat{\bm r}_{1})}{|\widehat{\bm \Omega} 
\times \widehat{\bm r}_{1}|}\,
\frac{\mathcal{P}_{\ell}^{1}(\widehat{\bm \Omega} \cdot \widehat{\bm r}_{2})}{|\widehat{\bm \Omega} 
\times \widehat{\bm r}_{2}|}\,{\widehat \Omega}^{a}\,{\widehat \Omega}^{b}\,.
\label{gamma_contract}
\end{align}
This is a symmetric matrix (i.e., $\Gamma^{ab}=\Gamma^{ba}$). For 
given unit vectors $\widehat{\bm r}_{1}$ and $\widehat{\bm r}_{2}$, the functional form of $\Gamma^{ab}$ is uniquely determined. Then, taking advantage of its rotational covariance, the $\Gamma^{ab}$ is generally expressed as follows:
\begin{align}
 \Gamma^{a b}(\ell, \mu) &=  F_{\ell}(\mu) \delta^{ab} 
+ G_{\ell}(\mu)({\widehat r}_{1}^{a} {\widehat r}_{2}^{b}+{\widehat r}_{2}^{a} {\widehat r}_{1}^{b})
\nonumber \\
&\quad+ H_{\ell}(\mu)({\widehat r}_{1}^{a} {\widehat r}_{1}^{b}+{\widehat r}_{2}^{a} {\widehat r}_{2}^{b}),
\label{gamma_sym}
\end{align}
with $\mu$ being the directional cosine, defined by $\mu\equiv \widehat{\bm r}_{1}\cdot\widehat{\bm r}_{2}$. The explicit form of the functions $F_\ell$, $G_\ell$, and $H_\ell$ will be derived below. Using Eq.~(\ref{gamma_sym}), the function $\gamma_\ell^{\rm B}$ given at Eq.~(\ref{gamma1}) is rewritten with
\begin{align}
&\gamma_{\ell}^{\rm B}(\widehat{\bm r}_{1},\widehat{\bm r}_{2}) = \frac{(2\ell+1)(\ell-1)!}{2\pi(\ell+1)!} 
\nonumber\\
&\qquad \times \Bigl[
F_\ell(\mu)\,\left\{\mu\,(\widehat{\bm X}_1 \cdot \widehat{\bm X}_2)
-(\widehat{\bm r}_{2} \cdot \widehat{\bm X}_1) \, (\widehat{\bm r}_{1} \cdot \widehat{\bm X}_2) \right\}
\nonumber \\
&\qquad
-G_\ell(\mu)\,\left\{(\widehat{\bm r}_1 \times \widehat{\bm r}_2) \cdot \widehat{\bm X}_1 \right\}
\, \left\{(\widehat{\bm r}_1 \times \widehat{\bm r}_2) \cdot \widehat{\bm X}_2 \right\}
\Bigr]\,.
\nonumber
\end{align}
This is Eq.~(\ref{eq:Gamma}).

Now, the task is to derive the expressions for $F_\ell$, $G_\ell$, and $H_\ell$. To do this, we contract Eq.~(\ref{gamma_sym}) with the three symmetric tensors, $\delta_{ab}\,, ({\widehat r}_{1\,a} {\widehat r}_{2\,b}+{\widehat r}_{1\,b} {\widehat r}_{2\,a})$ and $({\widehat r}_{1\,a} {\widehat r}_{1\,b}+{\widehat r}_{2\,a} {\widehat r}_{2\,b})$. We define
\begin{align}
&p_\ell(\mu):= \Gamma^{a b}(\ell, \mu) \, \delta_{ab}\,,
\label{eq:def_p_ell}
\\
&q_\ell(\mu):= \Gamma^{a b}(\ell, \mu) \, ({\widehat r}_{1\,a} {\widehat r}_{2\,b}+{\widehat r}_{1\,b} {\widehat r}_{2\,a})\,,
\label{eq:def_q_ell}
\\
&r_\ell(\mu):= \Gamma^{a b}(\ell, \mu) \, ({\widehat r}_{1\,a} {\widehat r}_{1\,b}+{\widehat r}_{2\,a} {\widehat r}_{2\,b})\,.
\label{eq:def_r_ell}
\end{align}
A straightforward calculation shows that
\begin{align}
\left(
\begin{array}{c}
p_\ell \\
\\
q_\ell \\
\\
r_\ell 
\end{array} 
\right)=\left(
\begin{array}{ccc}
3 & 2\mu & 2 \\
\\
2\mu & 2(1+\mu^{2}) & 4\mu\\
\\
2 & 4\mu & 2(1+\mu^{2})
\end{array}
\right) \left(
\begin{array}{c}
F_\ell \\
\\
G_\ell \\
\\
H_\ell 
\end{array}
\right),
\label{eq:matrix1}
\end{align}
which is inverted to give
\begin{align}
\left(
\begin{array}{c}
F_\ell \\
\\
G_\ell \\
\\
H_\ell
\end{array} 
\right)=\left(
\begin{array}{ccc}
1 & \frac{\mu}{1-\mu^2} & -\frac{1}{1-\mu^2} \\
\\
\frac{\mu}{1-\mu^2} & \frac{1+3\mu^2}{2(1-\mu^2)^2} & -\frac{2\mu}{(1-\mu^2)^2}\\
\\
-\frac{1}{1-\mu^2} & -\frac{2\mu}{(1-\mu^2)^2} & \frac{3+\mu^2}{2(1-\mu^2)^2}
\end{array}
\right) \left(
\begin{array}{c}
p_\ell \\
\\
q_\ell \\
\\
r_\ell 
\end{array}
\right)\,.
\label{eq:matrix2}
\end{align}
The derivation of the explicit expressions for $p_\ell$, $q_\ell$, and $r_\ell$ involves a direct evaluation of the integrals in Eqs.~(\ref{eq:def_p_ell})-(\ref{eq:def_r_ell}), which we will present in Appendix \ref{appendix:legendre}. Here, we just summarize the final results: 
\begin{align}
&p_\ell(\mu)=4\pi\,\frac{\mathcal{P}_\ell^1(\mu)}{\sqrt{1-\mu^2}}, 
\label{eq:p_analytic} 
\\
&q_\ell(\mu)=8\pi\,\left[ \frac{\mathcal{P}_{\ell+1}^1(\mu)}{\sqrt{1-\mu^2}}
-\frac{(\ell+1)(3\ell+1)}{2\ell+1}\,\mathcal{P}_\ell(\mu)\right],
\label{eq:q_analytic}
\\
&r_\ell(\mu)=8\pi\,\left[ \frac{\mathcal{P}_\ell^1(\mu)}{\sqrt{1-\mu^2}}
-\frac{\ell(\ell+1)}{2\ell+1}\,\mathcal{P}_{\ell-1}(\mu)\right]\,.
\label{eq:r_analytic}
\end{align}

Substituting these into Eq.~(\ref{eq:matrix2}), we obtain the analytic expressions for $F_\ell$ and $G_\ell$ necessary to compute the function $\gamma_\ell^{\rm B}$. Note that using the properties of (associated) Legendre polynomials, the resultant expressions can be further reduced, and in Eqs.~(\ref{eq:fl}) and (\ref{eq:gl}), we presented a simplified form of the functions $F_\ell$ and $G_\ell$.

\section{Explicit expressions for $p_\ell$, $q_\ell$, and $r_\ell$}
\label{appendix:legendre}

In this Appendix, computing the integrals given at Eqs.~(\ref{eq:def_p_ell})-(\ref{eq:def_r_ell}), we derive the analytic expressions for $p_\ell$, $q_\ell$, and $r_\ell$, summarized in Eqs.~(\ref{eq:p_analytic})-(\ref{eq:r_analytic}).

First notice that the integrands of these functions are expressed in terms of the quantities, $z_i$, defined by $z_i\equiv (\widehat{\bm \Omega}\cdot\widehat{\bm r}_i)$. Using the fact that $|\widehat{\bm \Omega}\times\widehat{\bm r}_i|=\sqrt{1-z_i^2}$, the integrals at Eqs.~(\ref{eq:def_p_ell})-(\ref{eq:def_r_ell}) are rewritten with
\begin{align}
&p_{\ell}=\int_{S^{2}}\,d^2{\widehat {\bm \Omega}}\,\,
\frac{d \mathcal{P}_{\ell}(z_{1})}{dz_{1}}\frac{d \mathcal{P}_{\ell}(z_{2})}{dz_{2}}\,, 
\label{eq:pl}
\\
&q_{\ell}=2\int_{S^{2}}\,d^2{\widehat {\bm \Omega}}\,\,
 \frac{d \mathcal{P}_{\ell}(z_{1})}{dz_{1}}  \frac{d \mathcal{P}_{\ell}(z_{2})}{dz_{2}}\,z_{1}z_{2},
 \label{eq:ql}
\\
&r_{\ell}=\int_{S^{2}}\,d^2{\widehat {\bm \Omega}}\,\,
\frac{d \mathcal{P}_{\ell}(z_{1})}{dz_{1}}\frac{d \mathcal{P}_{\ell}(z_{2})}{dz_{2}} ( z_{1}^{2} +z_{2}^{2}) 
\label{eq:rl}\,.
\end{align}
In deriving the expressions above, we used the relation between the associated and normal Legendre polynomials:
\begin{align}
\mathcal{P}_{\ell}^{1}(z)=\sqrt{1-z^{2}}\frac{d \mathcal{P}_{\ell}(z)}{dz}.
\label{eq:derivative}
\end{align}

In what follows, we will explicitly compute each integral with a help of mathematical formulas \cite{1960iwanami,2007tisp.book.....G}.

\subsection{Function $p_\ell$}
\label{subsec:func_p_ell}

Consider first Eq.~(\ref{eq:pl}). To analytically perform the integration, we wish to rewrite the integrand in a separable form. To do this, we use the following formula to express the derivative $d\mathcal{P}_\ell/dz$ in terms of the Legendre polynomials:
\begin{align}
\frac{d \mathcal{P}_{\ell}(z)}{dz}=
\left\{
 \begin{array}{ll} \displaystyle
\sum_{r=1}^{\ell/2}\,\;\;(4r-1)\,\mathcal{P}_{2r-1}(z), &\ell:{\rm even}\\
\displaystyle
\hspace{-2mm}\sum_{r=0}^{(\ell-1)/2}(4r+1)\,\mathcal{P}_{2r}(z), & \ell:{\rm odd} 
 \end{array}
 \right..
\label{eq:formula1}
\end{align}
For the cases with even $\ell$, Eq.~(\ref{eq:pl}) is rewritten with
\begin{align}
p_{\ell}=\sum_{r,r'=1}^{\ell/2}\,(4r-1)(4r'-1)\,\int_{S^{2}}\,d^2{\widehat {\bm \Omega}}\,\, \,\mathcal{P}_{2r-1}(z_1)\mathcal{P}_{2r'-1}(z_2).
\label{eq:p_ell_v2} 
\end{align}
We now apply the spherical harmonic addition theorem: 
\begin{align}
\mathcal{P}_{\ell}(\widehat{\bm \Omega} \cdot \widehat{\bm r}_{i})=\frac{4\pi}{2\ell+1}
\sum_{m=-\ell}^{\ell} Y_{\ell m}(\widehat{\bm  \Omega}) Y_{\ell m}^{\ast}(\widehat{\bm r}_{i}).
\label{eq:addition}
\end{align}
Eq.~(\ref{eq:p_ell_v2}) then leads to
\begin{align}
 p_{\ell}& =(4\pi)^2\sum_{r,r'=1}^{\ell/2}\sum_{m,m'}\,\,
\int_{S^{2}}\,d^2{\widehat {\bm \Omega}}\,\,
Y_{2r-1,m}(\widehat{\bm \Omega})\, Y_{2r-1,m}^*(\widehat{\bm r}_1)
\nonumber\\
& \qquad\qquad \qquad\qquad \qquad
\times\, Y_{2r'-1,m'}^*(\widehat{\bm \Omega})\, Y_{2r'-1,m'}(\widehat{\bm r}_2)
\nonumber\\
& = (4\pi)^2\sum_{r=1}^{\ell/2}\sum_{m}\,\,
Y_{2r-1,m}^*(\widehat{\bm r}_1)Y_{2r-1,m}(\widehat{\bm r}_2)
\nonumber\\
& = 4\pi\, \sum_{r=1}^{\ell/2}\,(4r-1)
\mathcal{P}_{2r-1}(\widehat{\bm r}_1\cdot\widehat{\bm r}_2).  
\label{eq:calculation_p_ell}
\end{align}
Here, in the second line, we have used the orthonormality of the spherical harmonics, $Y_{\ell m}$. In the last line, we again used the addition theorem. Then, 
from the formulas given at Eqs.~(\ref{eq:derivative}) and (\ref{eq:formula1}), we obtain
\begin{align}
p_\ell=4\pi \frac{\mathcal{P}_{\ell}^{1}(\widehat{\bm r}_1\cdot\widehat{\bm r}_2)}{\sqrt{1-(\widehat{\bm r}_1\cdot\widehat{\bm r}_2)^{2}}}\,.   
\nonumber
\end{align}
Denoting $\widehat{\bm r}_1\cdot\widehat{\bm r}_2$ by $\mu$, this is equivalent to Eq.~(\ref{eq:p_analytic}). While the above result was obtained for even $\ell$, it is shown that the final expression also holds for the cases with odd $\ell$.

\subsection{Function $q_\ell$}
\label{subsec:func_q_ell}

Consider next Eq.~(\ref{eq:ql}). To derive the analytic expression, we adopt the same approach as we did in Sec.~\ref{subsec:func_p_ell}, but instead of Eq.~(\ref{eq:formula1}), we use another formula:
\begin{align}
z\,\frac{d \mathcal{P}_{\ell}(z)}{dz}=
\left\{
 \begin{array}{ll} \displaystyle
\sum_{r=0}^{\ell/2}\;\;\,(4r+1)\,c_{2r}\,\mathcal{P}_{2r}(z), &\ell:{\rm even}\\
\displaystyle
\hspace{-2mm} \sum_{r=1}^{(\ell+1)/2}(4r-1)\,c_{2r-1}\,\mathcal{P}_{2r-1}(z), & \ell:{\rm odd} 
 \end{array}
 \right.,
\label{eq:formula2}
\end{align}
with the coefficient $c_m$ defined by (for $m=2r$ or $2r-1$)
\begin{align}
& c_{m}=\left\{
 \begin{array}{ll} \displaystyle
1\,, &(m<\ell) \\
\displaystyle
\frac{\ell}{2\ell+1}\,,& (m=\ell) 
 \end{array}
\label{eq:coeff_c_n}
 \right..
\end{align}
Eq.~(\ref{eq:formula2}) is derived from Eq.~(\ref{eq:formula1}), using the recursion relation, $(\ell+1)\mathcal{P}_{\ell+1}(z)-(2\ell+1)z\,\mathcal{P}_\ell(z)+\ell\,\mathcal{P}_{\ell-1}(z)=0$.

Substituting Eq.~(\ref{eq:formula2}) into Eq.~(\ref{eq:ql}), for even $\ell$, we obtain 
\begin{align}
& q_\ell=2 \sum_{r,r'=0}^{\ell/2}\,(4r+1)(4r'+1)\,c_{2r}\,c_{2r'} 
\nonumber\\
&\qquad\qquad \times
\int_{S^{2}}\,d^2{\widehat {\bm \Omega}}\,\, \,\mathcal{P}_{2r}(z_1)\mathcal{P}_{2r'}(z_2).
\label{eq:q_ell_v2} 
\end{align}
Then, applying the addition theorem at Eq.~(\ref{eq:addition}), we proceed to the calculations in similar way to Eq.~(\ref{eq:calculation_p_ell}). The result becomes (for even $\ell$)
\begin{align}
 q_\ell&=8\pi \sum_{r=0}^{\ell/2}\,(4r+1)\,c_{2r}^2\,\mathcal{P}_{2r}(\widehat{\bm r}_1\cdot\widehat{\bm r}_2)
\nonumber\\
&=8\pi \Biggl( \sum_{r=0}^{\ell/2}\,(4r+1)\,\mathcal{P}_{2r}(\widehat{\bm r}_1\cdot\widehat{\bm r}_2)
\nonumber\\
&\quad
+(c_\ell^2-1) (2\ell+1) \,\mathcal{P}_\ell(\widehat{\bm r}_1\cdot\widehat{\bm r}_2) \Biggr).
\label{eq:q_ell_v3} 
\end{align}
With a help of Eq.~(\ref{eq:coeff_c_n}), this is recast as
\begin{align}
 q_\ell &= 
8\pi\Biggl[\frac{\mathcal{P}_{\ell+1}^1(\widehat{\bm r}_1\cdot\widehat{\bm r}_2)}{\sqrt{1-(\widehat{\bm r}_1\cdot\widehat{\bm r}_2)^2}}-\frac{(\ell+1)(3\ell+1)}{2\ell+1}\,\mathcal{P}_\ell(\widehat{\bm r}_1\cdot\widehat{\bm r}_2)\Biggr]. 
\nonumber
\end{align}
This is Eq.~(\ref{eq:q_analytic}). Here, we used the formula given at Eq.~(\ref{eq:formula1}) and further rewrote the derivative $d\mathcal{P}_\ell/dz$ with $\mathcal{P}_\ell^1/\sqrt{1-z^{2}}$ through the relation in Eq.~(\ref{eq:derivative}). Note that the final expression is shown to also hold for odd $\ell$.

\subsection{Function $r_\ell$}
\label{subsec:func_r_ell}

Finally consider Eq.~(\ref{eq:rl}). We repeat almost the same calculations as we performed in Secs.~\ref{subsec:func_p_ell} and \ref{subsec:func_q_ell}, using the formula below:
\begin{align}
z^{2}\frac{d \mathcal{P}_{\ell}(z)}{dz}=
\left\{
 \begin{array}{ll} \displaystyle
\sum_{r=1}^{(\ell+2)/2}(4r-1)\, d_{2r-1}\, \mathcal{P}_{2r-1}(z), &\ell:{\rm even}\\
\displaystyle
\sum_{r=0}^{(\ell+1)/2}(4r+1)\,d_{2r}\,\mathcal{P}_{2r}(z), & \ell:{\rm odd} 
 \end{array}
 \right.,
\label{eq:formula3}
\end{align}
with the coefficient $d_{m}$ given by (for $m=2r-1$ or $2r$)\footnote{Eq.~(\ref{eq:formula3}) can be derived based on the formula at Eq.~(\ref{eq:formula1}), with a help of the following recursion formulas: 
\begin{align}
&(z^2-1)\frac{d\mathcal{P}_\ell(z)}{dz}=\ell\,\{z\,\mathcal{P}_\ell(z)-\mathcal{P}_{\ell-1}(z)\},
\nonumber\\
&(\ell+1)\mathcal{P}_{\ell+1}(z)-(2\ell+1)z\,\mathcal{P}_\ell(z)+\ell\,\mathcal{P}_{\ell-1}(z)=0.
\nonumber
\end{align}
}
\begin{align}
d_{m}=\left\{
 \begin{array}{ll} \displaystyle
1\,, &(m<\ell-1) \\
\displaystyle
1-\frac{\ell(\ell+1)}{(2\ell-1)(2\ell+1)}\,,& (m=\ell-1) \\
\displaystyle
\frac{\ell(\ell+1)}{(2\ell+1)(2\ell+3)}\,,& (m=\ell+1)
 \end{array}
 \right. .
\label{eq:coef_d_n}
\end{align}

Making use of the formula at Eq.~(\ref{eq:formula3}) and the addition theorem,  
Eq.~(\ref{eq:rl}) is recast as
\begin{align}
r_{\ell}&=
\left\{
 \begin{array}{ll} \displaystyle
8 \pi \sum_{r=1}^{\ell/2} \;\;\, (4r-1)\,d_{2r-1}\,\mathcal{P}_{2r-1}(\widehat{\bm r}_1\cdot\widehat{\bm r}_2), & \ell:{\rm even} \\
\displaystyle
8 \pi \!\!\!\! \sum_{r=0}^{(\ell-1)/2} (4r+1)\,d_{2r}\,\mathcal{P}_{2r}(\widehat{\bm r}_1\cdot\widehat{\bm r}_2), & \ell:{\rm odd}
 \end{array}
 \right. .
\end{align}
Again, with a help of Eq.~(\ref{eq:coef_d_n}) and the formula at Eq.~(\ref{eq:formula1}), the above expression is simplified, and we finally obtain
\begin{align}
r_\ell=8\pi \left[\frac{\mathcal{P}_{\ell}^{1}(\widehat{\bm r}_1\cdot\widehat{\bm r}_2)}{\sqrt{1-(\widehat{\bm r}_1\cdot\widehat{\bm r}_2)^{2}}}-\frac{\ell(\ell+1)}{2\ell+1}\mathcal{P}_{\ell-1}(\widehat{\bm r}_1\cdot\widehat{\bm r}_2)\right]\,. 
\nonumber
\end{align}
This is Eq.~(\ref{eq:r_analytic}).

\section{Asymptotic expressions of $\gamma_{\ell}^{\rm B}$}
\label{appendix:af}

In this Appendix, we derive the asymptotic expressions of $\gamma_\ell^{\rm B}$, given at Eqs.~(\ref{eq:gamma_beta0}) and (\ref{eq:gamma_B_antipodal}).

To derive the expressions, we first explicitly write down the functions $A_\ell$ and $B_\ell$ given at Eq.~(\ref{eq:A_ell}) and (\ref{eq:B_ell}). Using Eq.(\ref{eq:derivative}), we have
\begin{align}
A_{\ell}=-\frac{1}{\ell(\ell+1)}&\Biggl[(\ell+1)\{1-(\ell+1)\mu\}\mathcal{P}_{\ell+1}(\mu)
\nonumber
\\
&-\{ \ell+\mu-(1+\ell)\mu^{2} \} \frac{d \mathcal{P}_{\ell+1}(\mu)}{d\mu}\Biggr]\,, 
\label{eq:A_ell_explicit} 
\\
B_{\ell}=-\frac{1}{\ell(\ell+1)}&\Biggr[(\ell+1)\{1+(\ell+1)\mu\}\mathcal{P}_{\ell+1}(\mu)
\nonumber
\\
&+\{ \ell-\mu-(1+\ell)\mu^{2} \} \frac{d \mathcal{P}_{\ell+1}(\mu)}{d\mu}\Biggl]\,.
\label{eq:B_ell_explicit}
\end{align}
Note here that $\mu=\cos\beta$. Thus, the colocated ($\beta=0^\circ$) and antipodal ($\beta=180^\circ$) detectors imply $\mu\to1$ and $-1$, respectively. In these limits, the Legendre polynomials become
\begin{align} 
&\mathcal{P}_{\ell}(1)=1\,, \quad \mathcal{P}_{\ell}(-1)=(-1)^{\ell}\,. 
\nonumber 
\end{align}
Further, using the relation above, the formula given at Eq.(\ref{eq:formula1}) leads to
\begin{align} 
& \left.\frac{d \mathcal{P}_{\ell}(\mu)}{d\mu}\right|_{\mu=1}=\frac{\ell(\ell+1)}{2}\,,
\quad \left.\frac{d \mathcal{P}_{\ell}(\mu)}{d\mu}\right|_{\mu=-1}=\frac{\ell(\ell+1)}{2}(-1)^{\ell+1}\,.
\end{align}

Applying these formulas to Eqs.~(\ref{eq:A_ell_explicit}) and (\ref{eq:B_ell_explicit}), we immediately see that $A_\ell\to1$ and $B_\ell\to0$ for colocated detectors. On the other hand, we obtain $A_\ell=0$ and $B_\ell=(-1)^{\ell+1}$ for the antipodal detectors.


\begin{thebibliography}{36}%
\makeatletter
\providecommand \@ifxundefined [1]{%
 \@ifx{#1\undefined}
}%
\providecommand \@ifnum [1]{%
 \ifnum #1\expandafter \@firstoftwo
 \else \expandafter \@secondoftwo
 \fi
}%
\providecommand \@ifx [1]{%
 \ifx #1\expandafter \@firstoftwo
 \else \expandafter \@secondoftwo
 \fi
}%
\providecommand \natexlab [1]{#1}%
\providecommand \enquote  [1]{``#1''}%
\providecommand \bibnamefont  [1]{#1}%
\providecommand \bibfnamefont [1]{#1}%
\providecommand \citenamefont [1]{#1}%
\providecommand \href@noop [0]{\@secondoftwo}%
\providecommand \href [0]{\begingroup \@sanitize@url \@href}%
\providecommand \@href[1]{\@@startlink{#1}\@@href}%
\providecommand \@@href[1]{\endgroup#1\@@endlink}%
\providecommand \@sanitize@url [0]{\catcode `\\12\catcode `\$12\catcode
  `\&12\catcode `\#12\catcode `\^12\catcode `\_12\catcode `\%12\relax}%
\providecommand \@@startlink[1]{}%
\providecommand \@@endlink[0]{}%
\providecommand \url  [0]{\begingroup\@sanitize@url \@url }%
\providecommand \@url [1]{\endgroup\@href {#1}{\urlprefix }}%
\providecommand \urlprefix  [0]{URL }%
\providecommand \Eprint [0]{\href }%
\providecommand \doibase [0]{http://dx.doi.org/}%
\providecommand \selectlanguage [0]{\@gobble}%
\providecommand \bibinfo  [0]{\@secondoftwo}%
\providecommand \bibfield  [0]{\@secondoftwo}%
\providecommand \translation [1]{[#1]}%
\providecommand \BibitemOpen [0]{}%
\providecommand \bibitemStop [0]{}%
\providecommand \bibitemNoStop [0]{.\EOS\space}%
\providecommand \EOS [0]{\spacefactor3000\relax}%
\providecommand \BibitemShut  [1]{\csname bibitem#1\endcsname}%
\let\auto@bib@innerbib\@empty
\bibitem [{\citenamefont {Maggiore}(2000)}]{Maggiore:1999vm}%
  \BibitemOpen
  \bibfield  {author} {\bibinfo {author} {\bibfnamefont {M.}~\bibnamefont
  {Maggiore}},\ }\href {\doibase 10.1016/S0370-1573(99)00102-7} {\bibfield
  {journal} {\bibinfo  {journal} {Phys. Rept.}\ }\textbf {\bibinfo {volume}
  {331}},\ \bibinfo {pages} {283} (\bibinfo {year} {2000})},\ \Eprint
  {http://arxiv.org/abs/gr-qc/9909001} {arXiv:gr-qc/9909001 [gr-qc]}
  \BibitemShut {NoStop}%
\bibitem [{\citenamefont {Abbott}\ \emph
  {et~al.}(2016{\natexlab{a}})\citenamefont {Abbott} \emph
  {et~al.}}]{Abbott:2016blz}%
  \BibitemOpen
  \bibfield  {author} {\bibinfo {author} {\bibfnamefont {B.~P.}\ \bibnamefont
  {Abbott}} \emph {et~al.} (\bibinfo {collaboration} {Virgo, LIGO
  Scientific}),\ }\href {\doibase 10.1103/PhysRevLett.116.061102} {\bibfield
  {journal} {\bibinfo  {journal} {Phys. Rev. Lett.}\ }\textbf {\bibinfo
  {volume} {116}},\ \bibinfo {pages} {061102} (\bibinfo {year}
  {2016}{\natexlab{a}})},\ \Eprint {http://arxiv.org/abs/1602.03837}
  {arXiv:1602.03837 [gr-qc]} \BibitemShut {NoStop}%
\bibitem [{\citenamefont {Abbott}\ \emph
  {et~al.}(2016{\natexlab{b}})\citenamefont {Abbott} \emph
  {et~al.}}]{TheLIGOScientific:2016wyq}%
  \BibitemOpen
  \bibfield  {author} {\bibinfo {author} {\bibfnamefont {B.~P.}\ \bibnamefont
  {Abbott}} \emph {et~al.} (\bibinfo {collaboration} {Virgo, LIGO
  Scientific}),\ }\href {\doibase 10.1103/PhysRevLett.116.131102} {\bibfield
  {journal} {\bibinfo  {journal} {Phys. Rev. Lett.}\ }\textbf {\bibinfo
  {volume} {116}},\ \bibinfo {pages} {131102} (\bibinfo {year}
  {2016}{\natexlab{b}})},\ \Eprint {http://arxiv.org/abs/1602.03847}
  {arXiv:1602.03847 [gr-qc]} \BibitemShut {NoStop}%
\bibitem [{\citenamefont {Regimbau}(2011)}]{1674-4527-11-4-001}%
  \BibitemOpen
  \bibfield  {author} {\bibinfo {author} {\bibfnamefont {T.}~\bibnamefont
  {Regimbau}},\ }\href {http://stacks.iop.org/1674-4527/11/i=4/a=001}
  {\bibfield  {journal} {\bibinfo  {journal} {Research in Astronomy and
  Astrophysics}\ }\textbf {\bibinfo {volume} {11}},\ \bibinfo {pages} {369}
  (\bibinfo {year} {2011})}\BibitemShut {NoStop}%
\bibitem [{\citenamefont {Zhu}\ \emph {et~al.}(2011)\citenamefont {Zhu},
  \citenamefont {Howell}, \citenamefont {Regimbau}, \citenamefont {Blair},\
  and\ \citenamefont {Zhu}}]{Zhu:2011bd}%
  \BibitemOpen
  \bibfield  {author} {\bibinfo {author} {\bibfnamefont {X.-J.}\ \bibnamefont
  {Zhu}}, \bibinfo {author} {\bibfnamefont {E.}~\bibnamefont {Howell}},
  \bibinfo {author} {\bibfnamefont {T.}~\bibnamefont {Regimbau}}, \bibinfo
  {author} {\bibfnamefont {D.}~\bibnamefont {Blair}}, \ and\ \bibinfo {author}
  {\bibfnamefont {Z.-H.}\ \bibnamefont {Zhu}},\ }\href {\doibase
  10.1088/0004-637X/739/2/86} {\bibfield  {journal} {\bibinfo  {journal}
  {Astrophys. J.}\ }\textbf {\bibinfo {volume} {739}},\ \bibinfo {pages} {86}
  (\bibinfo {year} {2011})},\ \Eprint {http://arxiv.org/abs/1104.3565}
  {arXiv:1104.3565 [gr-qc]} \BibitemShut {NoStop}%
\bibitem [{\citenamefont {Rosado}(2011)}]{Rosado:2011kv}%
  \BibitemOpen
  \bibfield  {author} {\bibinfo {author} {\bibfnamefont {P.~A.}\ \bibnamefont
  {Rosado}},\ }\href {\doibase 10.1103/PhysRevD.84.084004} {\bibfield
  {journal} {\bibinfo  {journal} {Phys. Rev.}\ }\textbf {\bibinfo {volume}
  {D84}},\ \bibinfo {pages} {084004} (\bibinfo {year} {2011})},\ \Eprint
  {http://arxiv.org/abs/1106.5795} {arXiv:1106.5795 [gr-qc]} \BibitemShut
  {NoStop}%
\bibitem [{\citenamefont {Marassi}\ \emph {et~al.}(2011)\citenamefont
  {Marassi}, \citenamefont {Schneider}, \citenamefont {Corvino}, \citenamefont
  {Ferrari},\ and\ \citenamefont {Portegies~Zwart}}]{Marassi:2011si}%
  \BibitemOpen
  \bibfield  {author} {\bibinfo {author} {\bibfnamefont {S.}~\bibnamefont
  {Marassi}}, \bibinfo {author} {\bibfnamefont {R.}~\bibnamefont {Schneider}},
  \bibinfo {author} {\bibfnamefont {G.}~\bibnamefont {Corvino}}, \bibinfo
  {author} {\bibfnamefont {V.}~\bibnamefont {Ferrari}}, \ and\ \bibinfo
  {author} {\bibfnamefont {S.}~\bibnamefont {Portegies~Zwart}},\ }\href
  {\doibase 10.1103/PhysRevD.84.124037} {\bibfield  {journal} {\bibinfo
  {journal} {Phys. Rev.}\ }\textbf {\bibinfo {volume} {D84}},\ \bibinfo {pages}
  {124037} (\bibinfo {year} {2011})},\ \Eprint {http://arxiv.org/abs/1111.6125}
  {arXiv:1111.6125 [astro-ph.CO]} \BibitemShut {NoStop}%
\bibitem [{\citenamefont {Wu}\ \emph {et~al.}(2012)\citenamefont {Wu},
  \citenamefont {Mandic},\ and\ \citenamefont {Regimbau}}]{Wu:2011ac}%
  \BibitemOpen
  \bibfield  {author} {\bibinfo {author} {\bibfnamefont {C.}~\bibnamefont
  {Wu}}, \bibinfo {author} {\bibfnamefont {V.}~\bibnamefont {Mandic}}, \ and\
  \bibinfo {author} {\bibfnamefont {T.}~\bibnamefont {Regimbau}},\ }\href
  {\doibase 10.1103/PhysRevD.85.104024} {\bibfield  {journal} {\bibinfo
  {journal} {Phys. Rev.}\ }\textbf {\bibinfo {volume} {D85}},\ \bibinfo {pages}
  {104024} (\bibinfo {year} {2012})},\ \Eprint {http://arxiv.org/abs/1112.1898}
  {arXiv:1112.1898 [gr-qc]} \BibitemShut {NoStop}%
\bibitem [{\citenamefont {Wu}\ \emph {et~al.}(2013)\citenamefont {Wu},
  \citenamefont {Mandic},\ and\ \citenamefont {Regimbau}}]{Wu:2013xfa}%
  \BibitemOpen
  \bibfield  {author} {\bibinfo {author} {\bibfnamefont {C.-J.}\ \bibnamefont
  {Wu}}, \bibinfo {author} {\bibfnamefont {V.}~\bibnamefont {Mandic}}, \ and\
  \bibinfo {author} {\bibfnamefont {T.}~\bibnamefont {Regimbau}},\ }\href
  {\doibase 10.1103/PhysRevD.87.042002} {\bibfield  {journal} {\bibinfo
  {journal} {Phys. Rev.}\ }\textbf {\bibinfo {volume} {D87}},\ \bibinfo {pages}
  {042002} (\bibinfo {year} {2013})}\BibitemShut {NoStop}%
\bibitem [{\citenamefont {Mandic}\ \emph {et~al.}(2016)\citenamefont {Mandic},
  \citenamefont {Bird},\ and\ \citenamefont {Cholis}}]{Mandic:2016lcn}%
  \BibitemOpen
  \bibfield  {author} {\bibinfo {author} {\bibfnamefont {V.}~\bibnamefont
  {Mandic}}, \bibinfo {author} {\bibfnamefont {S.}~\bibnamefont {Bird}}, \ and\
  \bibinfo {author} {\bibfnamefont {I.}~\bibnamefont {Cholis}},\ }\href
  {\doibase 10.1103/PhysRevLett.117.201102} {\bibfield  {journal} {\bibinfo
  {journal} {Phys. Rev. Lett.}\ }\textbf {\bibinfo {volume} {117}},\ \bibinfo
  {pages} {201102} (\bibinfo {year} {2016})},\ \Eprint
  {http://arxiv.org/abs/1608.06699} {arXiv:1608.06699 [astro-ph.CO]}
  \BibitemShut {NoStop}%
\bibitem [{\citenamefont {Allen}\ and\ \citenamefont
  {Romano}(1999)}]{Allen:1997ad}%
  \BibitemOpen
  \bibfield  {author} {\bibinfo {author} {\bibfnamefont {B.}~\bibnamefont
  {Allen}}\ and\ \bibinfo {author} {\bibfnamefont {J.~D.}\ \bibnamefont
  {Romano}},\ }\href {\doibase 10.1103/PhysRevD.59.102001} {\bibfield
  {journal} {\bibinfo  {journal} {Phys.Rev.}\ }\textbf {\bibinfo {volume}
  {D59}},\ \bibinfo {pages} {102001} (\bibinfo {year} {1999})},\ \Eprint
  {http://arxiv.org/abs/gr-qc/9710117} {arXiv:gr-qc/9710117 [gr-qc]}
  \BibitemShut {NoStop}%
\bibitem [{\citenamefont {Christensen}(1992)}]{Christensen:1992wi}%
  \BibitemOpen
  \bibfield  {author} {\bibinfo {author} {\bibfnamefont {N.}~\bibnamefont
  {Christensen}},\ }\href {\doibase 10.1103/PhysRevD.46.5250} {\bibfield
  {journal} {\bibinfo  {journal} {Phys. Rev.}\ }\textbf {\bibinfo {volume}
  {D46}},\ \bibinfo {pages} {5250} (\bibinfo {year} {1992})}\BibitemShut
  {NoStop}%
\bibitem [{\citenamefont
  {{Schumann}}(1952{\natexlab{a}})}]{1952ZNatA...7..149S}%
  \BibitemOpen
  \bibfield  {author} {\bibinfo {author} {\bibfnamefont {W.~O.}\ \bibnamefont
  {{Schumann}}},\ }\href {\doibase 10.1515/zna-1952-0202} {\bibfield  {journal}
  {\bibinfo  {journal} {Zeitschrift Naturforschung Teil A}\ }\textbf {\bibinfo
  {volume} {7}},\ \bibinfo {pages} {149} (\bibinfo {year}
  {1952}{\natexlab{a}})}\BibitemShut {NoStop}%
\bibitem [{\citenamefont
  {{Schumann}}(1952{\natexlab{b}})}]{1952ZNatA...7..250S}%
  \BibitemOpen
  \bibfield  {author} {\bibinfo {author} {\bibfnamefont {W.~O.}\ \bibnamefont
  {{Schumann}}},\ }\href {\doibase 10.1515/zna-1952-3-404} {\bibfield
  {journal} {\bibinfo  {journal} {Zeitschrift Naturforschung Teil A}\ }\textbf
  {\bibinfo {volume} {7}},\ \bibinfo {pages} {250} (\bibinfo {year}
  {1952}{\natexlab{b}})}\BibitemShut {NoStop}%
\bibitem [{\citenamefont {Kowalska-Leszczynska}\ \emph
  {et~al.}(2017)\citenamefont {Kowalska-Leszczynska} \emph
  {et~al.}}]{Kowalska-Leszczynska:2016low}%
  \BibitemOpen
  \bibfield  {author} {\bibinfo {author} {\bibfnamefont {I.}~\bibnamefont
  {Kowalska-Leszczynska}} \emph {et~al.},\ }\href {\doibase
  10.1088/1361-6382/aa60eb} {\bibfield  {journal} {\bibinfo  {journal} {Class.
  Quant. Grav.}\ }\textbf {\bibinfo {volume} {34}},\ \bibinfo {pages} {074002}
  (\bibinfo {year} {2017})},\ \Eprint {http://arxiv.org/abs/1612.01102}
  {arXiv:1612.01102 [astro-ph.IM]} \BibitemShut {NoStop}%
\bibitem [{\citenamefont {Harry}(2010)}]{Harry:2010zz}%
  \BibitemOpen
  \bibfield  {author} {\bibinfo {author} {\bibfnamefont {G.~M.}\ \bibnamefont
  {Harry}} (\bibinfo {collaboration} {LIGO Scientific}),\ }\bibfield
  {booktitle} {\emph {\bibinfo {booktitle} {{Gravitational waves. Proceedings,
  8th Edoardo Amaldi Conference, Amaldi 8, New York, USA, June 22-26, 2009}}},\
  }\href {\doibase 10.1088/0264-9381/27/8/084006} {\bibfield  {journal}
  {\bibinfo  {journal} {Class. Quant. Grav.}\ }\textbf {\bibinfo {volume}
  {27}},\ \bibinfo {pages} {084006} (\bibinfo {year} {2010})}\BibitemShut
  {NoStop}%
\bibitem [{\citenamefont {Aasi}\ \emph {et~al.}(2015)\citenamefont {Aasi} \emph
  {et~al.}}]{TheLIGOScientific:2014jea}%
  \BibitemOpen
  \bibfield  {author} {\bibinfo {author} {\bibfnamefont {J.}~\bibnamefont
  {Aasi}} \emph {et~al.} (\bibinfo {collaboration} {LIGO Scientific}),\ }\href
  {\doibase 10.1088/0264-9381/32/7/074001} {\bibfield  {journal} {\bibinfo
  {journal} {Class. Quant. Grav.}\ }\textbf {\bibinfo {volume} {32}},\ \bibinfo
  {pages} {074001} (\bibinfo {year} {2015})},\ \Eprint
  {http://arxiv.org/abs/1411.4547} {arXiv:1411.4547 [gr-qc]} \BibitemShut
  {NoStop}%
\bibitem [{\citenamefont {Acernese}\ \emph {et~al.}(2015)\citenamefont
  {Acernese} \emph {et~al.}}]{TheVirgo:2014hva}%
  \BibitemOpen
  \bibfield  {author} {\bibinfo {author} {\bibfnamefont {F.}~\bibnamefont
  {Acernese}} \emph {et~al.} (\bibinfo {collaboration} {VIRGO}),\ }\href
  {\doibase 10.1088/0264-9381/32/2/024001} {\bibfield  {journal} {\bibinfo
  {journal} {Class. Quant. Grav.}\ }\textbf {\bibinfo {volume} {32}},\ \bibinfo
  {pages} {024001} (\bibinfo {year} {2015})},\ \Eprint
  {http://arxiv.org/abs/1408.3978} {arXiv:1408.3978 [gr-qc]} \BibitemShut
  {NoStop}%
\bibitem [{\citenamefont {{Thrane}}\ \emph {et~al.}(2013)\citenamefont
  {{Thrane}}, \citenamefont {{Christensen}},\ and\ \citenamefont
  {{Schofield}}}]{2013PhRvD..87l3009T}%
  \BibitemOpen
  \bibfield  {author} {\bibinfo {author} {\bibfnamefont {E.}~\bibnamefont
  {{Thrane}}}, \bibinfo {author} {\bibfnamefont {N.}~\bibnamefont
  {{Christensen}}}, \ and\ \bibinfo {author} {\bibfnamefont {R.~M.~S.}\
  \bibnamefont {{Schofield}}},\ }\href {\doibase 10.1103/PhysRevD.87.123009}
  {\bibfield  {journal} {\bibinfo  {journal} {\prd}\ }\textbf {\bibinfo
  {volume} {87}},\ \bibinfo {eid} {123009} (\bibinfo {year} {2013})},\ \Eprint
  {http://arxiv.org/abs/1303.2613} {arXiv:1303.2613 [astro-ph.IM]} \BibitemShut
  {NoStop}%
\bibitem [{\citenamefont {{Thrane}}\ \emph {et~al.}(2014)\citenamefont
  {{Thrane}}, \citenamefont {{Christensen}}, \citenamefont {{Schofield}},\ and\
  \citenamefont {{Effler}}}]{2014PhRvD..90b3013T}%
  \BibitemOpen
  \bibfield  {author} {\bibinfo {author} {\bibfnamefont {E.}~\bibnamefont
  {{Thrane}}}, \bibinfo {author} {\bibfnamefont {N.}~\bibnamefont
  {{Christensen}}}, \bibinfo {author} {\bibfnamefont {R.~M.~S.}\ \bibnamefont
  {{Schofield}}}, \ and\ \bibinfo {author} {\bibfnamefont {A.}~\bibnamefont
  {{Effler}}},\ }\href {\doibase 10.1103/PhysRevD.90.023013} {\bibfield
  {journal} {\bibinfo  {journal} {Physical Review D}\ }\textbf {\bibinfo
  {volume} {90}},\ \bibinfo {eid} {023013} (\bibinfo {year} {2014})},\ \Eprint
  {http://arxiv.org/abs/1406.2367} {arXiv:1406.2367 [astro-ph.IM]} \BibitemShut
  {NoStop}%
\bibitem [{\citenamefont {Coughlin}\ \emph {et~al.}(2016)\citenamefont
  {Coughlin} \emph {et~al.}}]{Coughlin:2016vor}%
  \BibitemOpen
  \bibfield  {author} {\bibinfo {author} {\bibfnamefont {M.~W.}\ \bibnamefont
  {Coughlin}} \emph {et~al.},\ }\href {\doibase 10.1088/0264-9381/33/22/224003}
  {\bibfield  {journal} {\bibinfo  {journal} {Class. Quant. Grav.}\ }\textbf
  {\bibinfo {volume} {33}},\ \bibinfo {pages} {224003} (\bibinfo {year}
  {2016})},\ \Eprint {http://arxiv.org/abs/1606.01011} {arXiv:1606.01011
  [gr-qc]} \BibitemShut {NoStop}%
\bibitem [{\citenamefont {Somiya}(2012)}]{Somiya:2011np}%
  \BibitemOpen
  \bibfield  {author} {\bibinfo {author} {\bibfnamefont {K.}~\bibnamefont
  {Somiya}} (\bibinfo {collaboration} {KAGRA}),\ }\bibfield  {booktitle} {\emph
  {\bibinfo {booktitle} {{Gravitational waves. Numerical relativity - data
  analysis. Proceedings, 9th Edoardo Amaldi Conference, Amaldi 9, and meeting,
  NRDA 2011, Cardiff, UK, July 10-15, 2011}}},\ }\href {\doibase
  10.1088/0264-9381/29/12/124007} {\bibfield  {journal} {\bibinfo  {journal}
  {Class. Quant. Grav.}\ }\textbf {\bibinfo {volume} {29}},\ \bibinfo {pages}
  {124007} (\bibinfo {year} {2012})},\ \Eprint {http://arxiv.org/abs/1111.7185}
  {arXiv:1111.7185 [gr-qc]} \BibitemShut {NoStop}%
\bibitem [{\citenamefont {Unnikrishnan}(2013)}]{Unnikrishnan:2013qwa}%
  \BibitemOpen
  \bibfield  {author} {\bibinfo {author} {\bibfnamefont {C.~S.}\ \bibnamefont
  {Unnikrishnan}},\ }\href {\doibase 10.1142/S0218271813410101} {\bibfield
  {journal} {\bibinfo  {journal} {Int. J. Mod. Phys.}\ }\textbf {\bibinfo
  {volume} {D22}},\ \bibinfo {pages} {1341010} (\bibinfo {year} {2013})},\
  \Eprint {http://arxiv.org/abs/1510.06059} {arXiv:1510.06059
  [physics.ins-det]} \BibitemShut {NoStop}%
\bibitem [{\citenamefont {Flanagan}(1993)}]{Flanagan:1993ix}%
  \BibitemOpen
  \bibfield  {author} {\bibinfo {author} {\bibfnamefont {E.~E.}\ \bibnamefont
  {Flanagan}},\ }\href {\doibase 10.1103/PhysRevD.48.2389} {\bibfield
  {journal} {\bibinfo  {journal} {Phys.Rev.}\ }\textbf {\bibinfo {volume}
  {D48}},\ \bibinfo {pages} {2389} (\bibinfo {year} {1993})},\ \Eprint
  {http://arxiv.org/abs/astro-ph/9305029} {arXiv:astro-ph/9305029 [astro-ph]}
  \BibitemShut {NoStop}%
\bibitem [{\citenamefont {{Jackson}}(1998)}]{1998clel.book.....J}%
  \BibitemOpen
  \bibfield  {author} {\bibinfo {author} {\bibfnamefont {J.~D.}\ \bibnamefont
  {{Jackson}}},\ }\href@noop {} {\emph {\bibinfo {title} {Classical
  Electrodynamics, 3rd Edition, by John David Jackson, pp.~832.~ISBN
  0-471-30932-X.~Wiley-VCH , July 1998.}}}\ (\bibinfo  {publisher} {Wiley},\
  \bibinfo {year} {1998})\ p.\ \bibinfo {pages} {832}\BibitemShut {NoStop}%
\bibitem [{\citenamefont {Seto}\ and\ \citenamefont
  {Taruya}(2008)}]{Seto:2008sr}%
  \BibitemOpen
  \bibfield  {author} {\bibinfo {author} {\bibfnamefont {N.}~\bibnamefont
  {Seto}}\ and\ \bibinfo {author} {\bibfnamefont {A.}~\bibnamefont {Taruya}},\
  }\href {\doibase 10.1103/PhysRevD.77.103001} {\bibfield  {journal} {\bibinfo
  {journal} {Phys.Rev.}\ }\textbf {\bibinfo {volume} {D77}},\ \bibinfo {pages}
  {103001} (\bibinfo {year} {2008})},\ \Eprint {http://arxiv.org/abs/0801.4185}
  {arXiv:0801.4185 [astro-ph]} \BibitemShut {NoStop}%
\bibitem [{\citenamefont {{Sentman}}(1983)}]{1983JATP...45...55S}%
  \BibitemOpen
  \bibfield  {author} {\bibinfo {author} {\bibfnamefont {D.~D.}\ \bibnamefont
  {{Sentman}}},\ }\href {\doibase 10.1016/S0021-9169(83)80008-7} {\bibfield
  {journal} {\bibinfo  {journal} {Journal of Atmospheric and Terrestrial
  Physics}\ }\textbf {\bibinfo {volume} {45}},\ \bibinfo {pages} {55} (\bibinfo
  {year} {1983})}\BibitemShut {NoStop}%
\bibitem [{\citenamefont {{Price}}\ \emph {et~al.}(2007)\citenamefont
  {{Price}}, \citenamefont {{Pechony}},\ and\ \citenamefont
  {{Greenberg}}}]{2007quality}%
  \BibitemOpen
  \bibfield  {author} {\bibinfo {author} {\bibfnamefont {C.}~\bibnamefont
  {{Price}}}, \bibinfo {author} {\bibfnamefont {O.}~\bibnamefont {{Pechony}}},
  \ and\ \bibinfo {author} {\bibfnamefont {E.}~\bibnamefont {{Greenberg}}},\
  }\href@noop {} {\bibfield  {journal} {\bibinfo  {journal} {Journal of
  Lightning Research}\ }\textbf {\bibinfo {volume} {1}},\ \bibinfo {pages} {1}
  (\bibinfo {year} {2007})}\BibitemShut {NoStop}%
\bibitem [{\citenamefont {{Horton}}(2014)}]{Ryan:2014}%
  \BibitemOpen
  \bibfield  {author} {\bibinfo {author} {\bibfnamefont {R.}~\bibnamefont
  {{Horton}}},\ }\href {https://dcc.ligo.org/LIGO-G1400943/public} {\bibfield
  {journal} {\bibinfo  {journal} {https://dcc.ligo.org/LIGO-G1400943/public}\ }
  (\bibinfo {year} {2014})}\BibitemShut {NoStop}%
\bibitem [{\citenamefont {{Christensen}}(1990)}]{1990PhDT........90C}%
  \BibitemOpen
  \bibfield  {author} {\bibinfo {author} {\bibfnamefont {N.~L.}\ \bibnamefont
  {{Christensen}}, \bibfnamefont {Jr.}},\ }\emph {\bibinfo {title} {{On
  Measuring the Stochastic Gravitational Radiation Background with Laser
  Interferometric Antennas.}}},\ \href@noop {} {Ph.D. thesis},\ \bibinfo
  {school} {MASSACHUSETTS INSTITUTE OF TECHNOLOGY.} (\bibinfo {year}
  {1990})\BibitemShut {NoStop}%
\bibitem [{\citenamefont {{Shoemaker}}(2010)}]{Shoemaker:2014}%
  \BibitemOpen
  \bibfield  {author} {\bibinfo {author} {\bibfnamefont {D.}~\bibnamefont
  {{Shoemaker}}},\ }\href {https://dcc.ligo.org/LIGO-T0900288-v3/public}
  {\bibfield  {journal} {\bibinfo  {journal}
  {https://dcc.ligo.org/LIGO-T0900288-v3/public}\ } (\bibinfo {year}
  {2010})}\BibitemShut {NoStop}%
\bibitem [{\citenamefont {Manzotti}\ and\ \citenamefont
  {Dietz}(2012)}]{Manzotti:2012uw}%
  \BibitemOpen
  \bibfield  {author} {\bibinfo {author} {\bibfnamefont {A.}~\bibnamefont
  {Manzotti}}\ and\ \bibinfo {author} {\bibfnamefont {A.}~\bibnamefont
  {Dietz}},\ }\href@noop {} {\  (\bibinfo {year} {2012})},\ \Eprint
  {http://arxiv.org/abs/1202.4031} {arXiv:1202.4031 [gr-qc]} \BibitemShut
  {NoStop}%
\bibitem [{\citenamefont {{Effler}}(2015)}]{Effler:2015p}%
  \BibitemOpen
  \bibfield  {author} {\bibinfo {author} {\bibfnamefont {A.}~\bibnamefont
  {{Effler}}},\ }\href
  {https://alog.ligo-la.caltech.edu/aLOG/index.php?callRep=17851} {\bibfield
  {journal} {\bibinfo  {journal}
  {https://alog.ligo-la.caltech.edu/aLOG/index.php?callRep=17851}\ } (\bibinfo
  {year} {2015})}\BibitemShut {NoStop}%
\bibitem [{\citenamefont {Schutz}(2011)}]{Schutz:2011tw}%
  \BibitemOpen
  \bibfield  {author} {\bibinfo {author} {\bibfnamefont {B.~F.}\ \bibnamefont
  {Schutz}},\ }\href {\doibase 10.1088/0264-9381/28/12/125023} {\bibfield
  {journal} {\bibinfo  {journal} {Class. Quant. Grav.}\ }\textbf {\bibinfo
  {volume} {28}},\ \bibinfo {pages} {125023} (\bibinfo {year} {2011})},\
  \Eprint {http://arxiv.org/abs/1102.5421} {arXiv:1102.5421 [astro-ph.IM]}
  \BibitemShut {NoStop}%
\bibitem [{\citenamefont {{Moriguchi}}\ \emph {et~al.}(1960)\citenamefont
  {{Moriguchi}}, \citenamefont {{Udagawa}},\ and\ \citenamefont
  {{Hitotsumatsu}}}]{1960iwanami}%
  \BibitemOpen
  \bibfield  {author} {\bibinfo {author} {\bibfnamefont {S.}~\bibnamefont
  {{Moriguchi}}}, \bibinfo {author} {\bibfnamefont {K.}~\bibnamefont
  {{Udagawa}}}, \ and\ \bibinfo {author} {\bibfnamefont {S.}~\bibnamefont
  {{Hitotsumatsu}}},\ }\href@noop {} {\emph {\bibinfo {title} {{Mathematical
  formulas, III}}}}\ (\bibinfo  {publisher} {Iwanami [in Japanese]},\ \bibinfo
  {year} {1960})\BibitemShut {NoStop}%
\bibitem [{\citenamefont {{Gradshteyn}}\ \emph {et~al.}(2014)\citenamefont
  {{Gradshteyn}}, \citenamefont {{Ryzhik}}, \citenamefont {{Jeffrey}},\ and\
  \citenamefont {{Zwillinger}}}]{2007tisp.book.....G}%
  \BibitemOpen
  \bibfield  {author} {\bibinfo {author} {\bibfnamefont {I.~S.}\ \bibnamefont
  {{Gradshteyn}}}, \bibinfo {author} {\bibfnamefont {I.~M.}\ \bibnamefont
  {{Ryzhik}}}, \bibinfo {author} {\bibfnamefont {A.}~\bibnamefont {{Jeffrey}}},
  \ and\ \bibinfo {author} {\bibfnamefont {D.}~\bibnamefont {{Zwillinger}}},\
  }\href@noop {} {\emph {\bibinfo {title} {Table of Integrals, Series, and
  Products, Seventh Edition by I.~S.~Gradshteyn, I.~M.~Ryzhik, Alan Jeffrey,
  and Daniel Zwillinger.~Elsevier Academic Press, 2007.~ISBN 012-373637-4}}}\
  (\bibinfo  {publisher} {Academic Press},\ \bibinfo {year} {2014})\BibitemShut
  {NoStop}%
\end{thebibliography}
\bibliographystyle{apsrev4-1}
%


\end{document}